\documentclass[aps,prl,twocolumn,superscriptaddress]{revtex4-1}
\pdfoutput=1
\usepackage{graphicx}
\usepackage{subfigure}
\usepackage{amssymb}
\usepackage{amsmath}
\usepackage{empheq}
\usepackage{mathrsfs}
\usepackage{stmaryrd}
\usepackage[cyr]{aeguill}
\usepackage[english]{babel}
\usepackage[linktocpage]{hyperref}
\usepackage{tocvsec2}
\usepackage{color}
\usepackage{version}
\usepackage{textcomp}
\usepackage{textcomp}
\usepackage{gensymb}
\usepackage{caption}

\begin{document}

\title{Absence of Fermi surface reconstruction in pressure-driven overdoped YBCO}

\author{Stanley W.  Tozer}
\email{stan.tozer@magnet.fsu.edu}
\affiliation{National High Magnetic Field Laboratory (NHMFL),Florida State University, Tallahassee, Florida}
\author{William A. Coniglio}
\affiliation{National High Magnetic Field Laboratory (NHMFL),Florida State University, Tallahassee, Florida}
\author{Tobias F\"orster}
\affiliation{Hochfeld-Magnetlabor Dresden (HLD-EMFL),  Helmholtz-Zentrum Dresden-Rossendorf, 01328, Dresden, Germany}
\author{Doug A. Bonn}
\affiliation{Department of Physics and Astronomy, University of British Columbia, Vancouver, British Columbia V6T 1Z4, Canada}
\author{Walter N. Hardy}
\affiliation{Department of Physics and Astronomy, University of British Columbia, Vancouver, British Columbia V6T 1Z4, Canada}
\author{Ruixing Liang}
\affiliation{Department of Physics and Astronomy, University of British Columbia, Vancouver, British Columbia V6T 1Z4, Canada}
\author{Erik Kampert}
\affiliation{Hochfeld-Magnetlabor Dresden (HLD-EMFL), Helmholtz-Zentrum Dresden-Rossendorf, 01328, Dresden, Germany}
\affiliation{Institute of Fundamentals of Electrical Engineering, Helmut Schmidt University, 22043 Hamburg, Germany}
\author{Audrey D.  Grockowiak}
\affiliation{National High Magnetic Field Laboratory (NHMFL),Florida State University, Tallahassee, Florida}
\affiliation{Brazilian Synchrotron Light Laboratory (LNLS/Sirius), Brazilian Center for Research in Energy and Materials (CNPEM), Campinas, Brazil}
\affiliation{Leibniz-Institut für Festkörper- und Werkstoffforschung Dresden,  01069 Dresden, Germany}
\email{a.grockowiak@ifw-dresden.de}

\date{\today}

\begin{abstract}

The evolution of the critical superconducting temperature and field,  quantum oscillation frequencies  and effective  mass  $m^{*}$ in underdoped YBa$_2$Cu$_3$O$_{7-\delta}$ (YBCO) crystals ($p$ = 0.11, with $p$ the hole concentration per Cu atom) points to a partial suppression of the charge orders with increasing pressure up to 7 GPa,  mimicking doping.   Application of pressures up to 25 GPa pushes the sample to the overdoped side of the superconducting dome.  Contrary to other cuprates,  or to doping studies on YBCO,  the frequency of the quantum oscillations measured in that pressure range  do not support the picture of a Fermi-surface reconstruction in the overdoped regime, but possibly point to the existence of a new charge order.

\end{abstract}

\maketitle

\section{Introduction}
The pnictides, cuprates, and molecular conductors exhibit similar features, leading to the possibility of a common mechanism anchored in a universal phase diagram. Typical ingredients for such phase diagrams include an antiferromagnetic phase, one or more superconducting domes, and possibly one, or several quantum critical points (QCP). Chemical doping is a traditional means to look at such phenomena. 
\newline The temperature-oxygen doping phase diagram of YBa$_2$Cu$_3$O$_{7-\delta}$ (YBCO) is extremely rich and exhibits a narrow antiferromagnetic region at the lowest doping.  At higher doping,  a superconducting dome develops \cite{Wu1987} with a hole doping of $p$ = 0.05 at the onset. The dome extends up to $p$ = 0.17,  which corresponds to optimal doping,  past which no samples are available. A shoulder at $p$ = 0.08 is associated with a Fermi surface reconstruction (FSR) \cite{Taillefer2009} as indicated by a change in sign of the Hall \cite{LeBoeuf2007},\cite{LeBoeuf2011} and Seebeck coefficients \cite{Chang2010} from positive to negative at low temperatures,  as well as at high magnetic fields.  This change in sign has been attributed, by utilizing nuclear magnetic resonance (NMR)  \cite{Wu2011} and x-ray diffraction (XRD) \cite{Chang2012},  to the presence of a two-dimensional charge order with a maximum around $p$ = 0.12.  Subsequently,  a field induced three-dimensional charge order (CO) was revealed by XRD   \cite{Gerber2015}, ultrasound \cite{Laliberte2018},  and NMR measurements \cite{Wu2013} in the same doping range. This has been recreated in zero field by the application of uniaxial strain along the $a$-axis of its orthorhombic structure \cite{Kim2018}. 
\\It is widely accepted that the Fermi surface (FS) undergoes a second reconstruction with oxygen doping, small electron and hole pockets evolving to arcs \cite{Doiron-Leyraud2007} and finally to a large pocket on the overdoped side \cite{Hussey2003},\cite{Vignolle2008}.  This picture is drawn from looking at the charge-carrier concentration in the copper-oxygen planes, and by comparing various cuprates at different doping levels   with Tl$_2$Ba$_2$CuO$_{6+\delta}$, the overdoped analogue to YBCO \cite{Barisic2013}. This is further supported by angle-resolved photoemission spectroscopy (ARPES) measurements of underdoped YBa$_2$Cu$_3$O$_{6.5}$ (YBCO6.5) crystals. These studies \cite{Hossain2008} have shown that the surface is naturally overdoped and different from the bulk, having found large Fermi arcs instead of the pockets measured via Shubnikov-de Haas (SdH) \cite{Doiron-Leyraud2007}.  By depositing K$^{+}$ ions on the surface, the authors reduced the charge-carrier doping at the surface and recovered the small Fermi-surface pockets \cite{Hossain2008}.  The only study showing a Fermi surface reconstruction (FSR), from underdoped to overdoped within one material, was carried out on the electron-doped Nd$_{2-x}$Ce$_{x}$CuO$_{4}$ compound via SdH measurements \cite{Helm2009}. \\
\newline
Additionally,  fermiology studies of various oxygen dopings in YBCO have revealed a divergence of the effective mass of the charge carriers close to optimal doping, hinting at the existence of a quantum critical point (QCP) under the superconducting dome \cite{Ramshaw317}. The relationship between this putative QCP and the FSR is not completely established.  \\

Both the QCP and the FSR  are crucial to our understanding of the cuprates as well as the universal phase diagram. 
A study of the Fermi surface via quantum oscillations (QO), with the aim of observing a reconstruction,  requires suppressing the superconducting phase to look at the normal state of the material close to optimal doping, and beyond.  Ideally, high magnetic fields are used to suppress superconductivity allowing for the observation of QO in the vicinity of the QCP; however, due to the high superconducting critical transition temperature ($T_{c}$) and concomitant superconducting critical field ($H_{c2}$),  this would require fields in excess of 100 T \cite{Grissonnanche2014}, beyond that currently available in non-destructive magnets.  Instead, doping with Zn up to  $p$=0.22 has been used to suppress the superconducting dome to a maximum $T_{c}$ of about 30 K \cite{Tallon1997},  but doping to these levels  introduces impurities and thus scattering which precludes the observation of QO. \\
To date,  high pressure studies look mostly at the evolution of $T_{c}$ with pressure \cite{Alireza2017},  \cite{Sadewasser1997},  \cite{Tozer1987}, \cite{Tozer1993}.   Here,  we have used quasi-hydrostatic pressure to tune the FS of YBCO6.5 starting from the underdoped region, $p$=0.11,  and progressing through the superconducting dome.  Measuring the evolution of QO with pressure and temperature provided direct access to the FSs and the effective masses of the charge carriers, which showed a lack of Fermi surface reconstruction,  and possibly points instead to the existence of a charge order similar to the that found on the underdoped side.\\

\section{Experimental methods}
We have used high purity twinned single crystals of YBCO6.5 ($p$=0.11) to perform high magnetic field Shubnikov-de Haas measurements under high pressure. Our experiments were carried out between 350 mK and 10 K in DC fields up to 45 T and pressures up to 25 GPa at the NHMFL using metallic diamond anvil cells (DACs),  and in pulsed fields up to 80 T and 7 GPa at HDL-EMFL using plastic DACs \cite{Graf2011},  rotators, and cryogenic tails.  The sample is coupled to a resonant LC tank circuit driven by a tunnel diode oscillator (TDO).  The small coil that makes up the inductor of this LC circuit resides in the high pressure volume of the DAC and senses changes in sample resistivity due to variations in temperature, pressure, or magnetic field (see supplementary information). This allowed us to develop a field-pressure-temperature phase diagram.  All data is taken with the $c$-axis of the crystal oriented parallel to the magnetic field.

\section{Results}
\subsection{Pressure effect on $T_c$ in zero field}

\begin{figure}[]
\begin{center}	
	\includegraphics[scale=0.29]{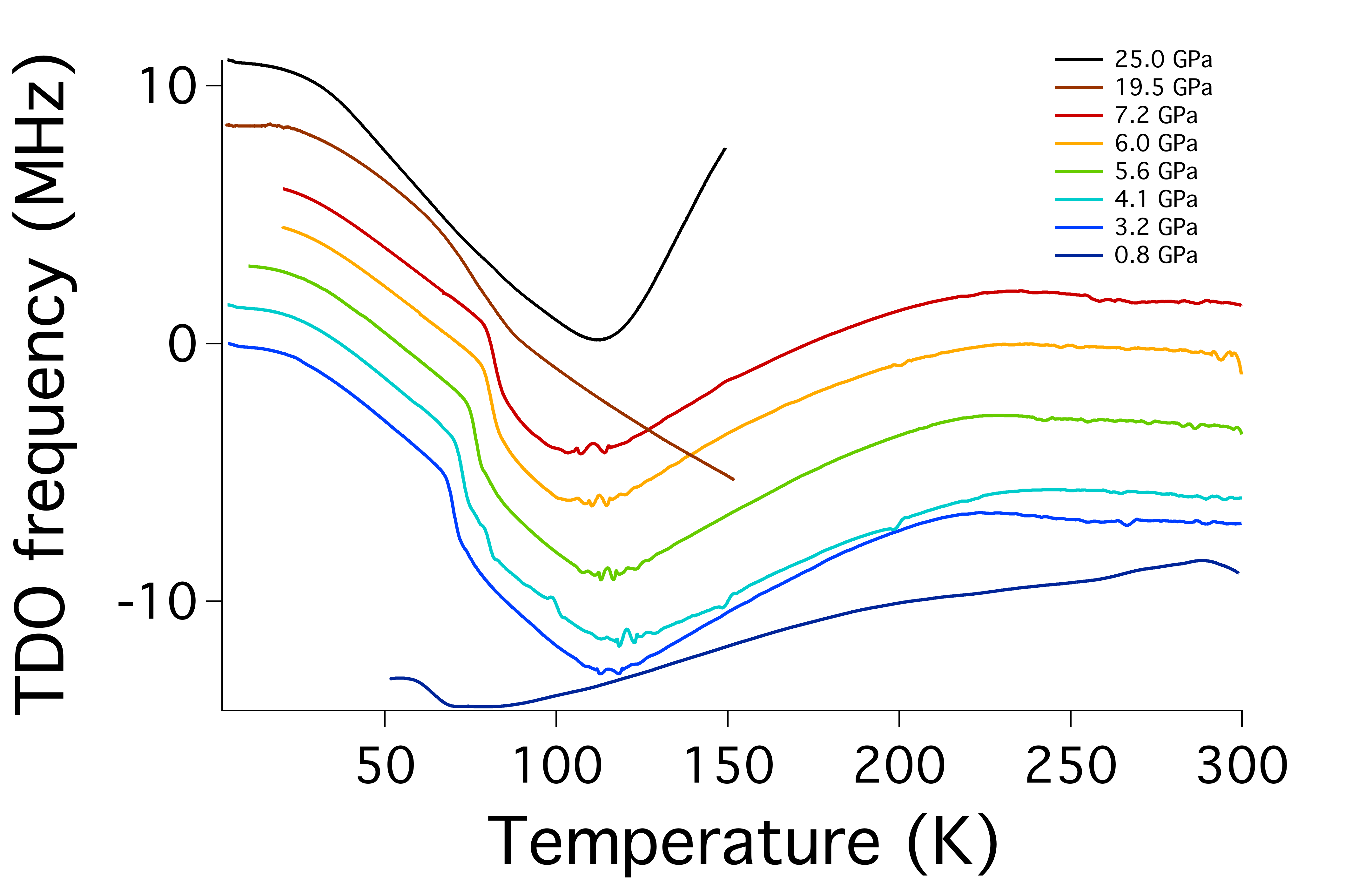}
		\caption{Evolution of the TDO frequency as a function of temperature for several pressures at 0 T.  The kink around 80 K is interpreted as the superconducting transition,  and $T_{c}$ is taken as the minimum in the derivative of the signal in that temperature range as shown in Sup.  Fig. \ref{Tc_diff.fig}. The observed kinks in those curves cannot be interpreted as $T_{c}$ as they are not suppressed with field as shown in Sup.  Fig. \ref{SI5_Hc_T_dep.fig}).  No clear transition appears at 25 GPa.  The minima observed in the signals at 115 K is an experimental artifact due to cross-over of frequency mixing. }
		\label{Tcvsp.fig}	
\end{center}		
\end{figure}

In our study,  the pressure was changed at room temperature, and was only increased during the course of this work,  i.e.  no measurements upon decompression were performed. The large vertical thermal gradient existing in cryostats upon loading a probe in DC fields or pulsed magnets makes any accurate determination of $T_c$ difficult and time constrains did not allow for controlled warming of the sample during high field time.  Careful $T_c$ measurements were performed in a $^{4}$He,  16 T Quantum Design PPMS, with the pressure being measured at the transition temperature.  Figure \ref{Tcvsp.fig} shows the evolution of the TDO frequency with temperature for various pressures. We define $T_c$ as the minimum of the derivative of the signal around 80 K for the pressures up to 7.2 GPa,  over which range $T_c$ smoothly increases up 81.2K.  At 19.5 GPa,  $T_c$ decreases to 75 K,  pointing to the sample having been driven to the overdoped side of the superconducting dome.  At 25 GPa,  we do not observe a similar kink down to 2 K in the PPMS.  The minimum around 100 K in the 25 GPa cell in Fig. \ref{Tcvsp.fig} cannot be interpreted as a $T_c$, as applied fields of 16 T did not shift it to lower temperatures (Sup.  Fig. \ref{SI4_Cell_25GPa.fig}).  Rather a combination of non-hydrostaticity,  and pressure variation upon cooling can explain the lack of $T_c$ at that pressure., both of which tend to broaden transitions  which make $T_c$ more difficult to measure. The cells used at higher pressure also have a larger variation in pressure with temperature: we measured an increase in pressure of roughly 20$\%$ between room temperature and $^{4}$He temperatures.

\subsection{ Pressure effect on the critical field}
\subsubsection{Pressures lower than 10 GPa}

\begin{figure}[]
\begin{center}	
	\includegraphics[scale=0.29]{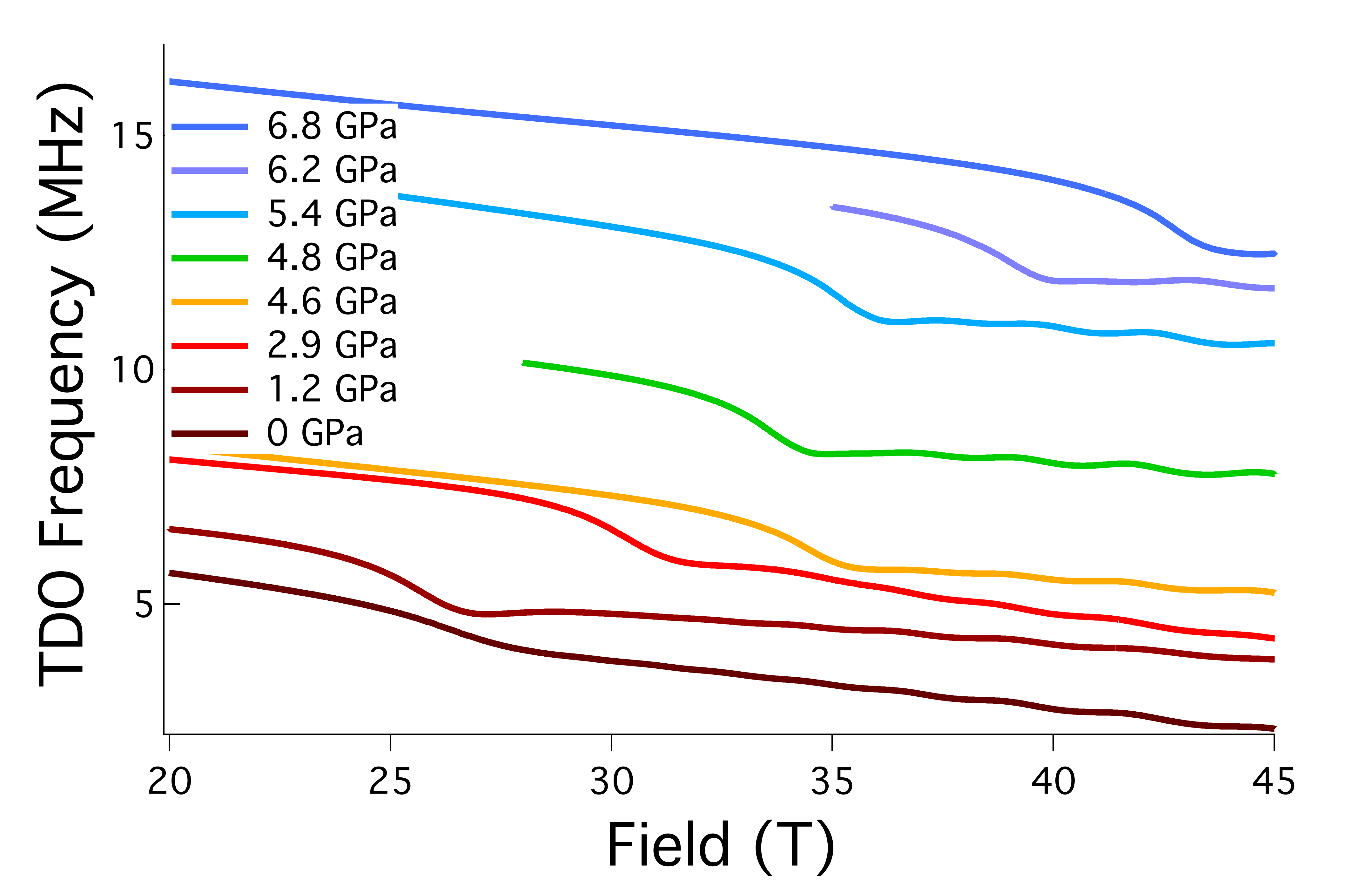}
		\caption{TDO frequencies as a function of magnetic field for several pressures at 400 mK.  The TDO frequencies are vertically shifted for clarity.  QO are clearly visible above the critical field. }
		\label{Frequency_vs_Field_hybrid.fig}	
\end{center}		
\end{figure}
The TDO frequencies as a function of the magnetic field measured in the 45 T hybrid magnet are plotted in Fig. \ref{Frequency_vs_Field_hybrid.fig} for various pressures at approximately 400 mK.  A clear trend in $H_{c2}$ appears,  dramatically increasing in a non-monotonic fashion up to 6.8 GPa,  where $H_{c2}$ is about 44 T.  The 45 T hybrid magnet limits our QO study to about 6 GPa for the underdoped side of the superconducting dome of YBCO.  Pulsed magnetic fields are required to measure the evolution of $H_{c2}$ and the QO frequencies for the higher pressure range.

\begin{figure}[]
\begin{center}	
	\includegraphics[scale=0.53]{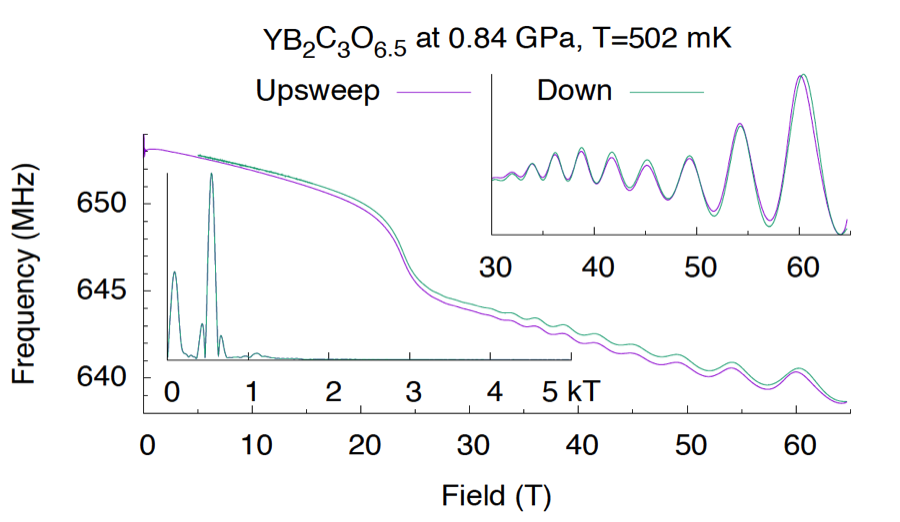}
		\caption{Main panel: TDO frequency as a function of magnetic field for a sample at 502 mK and 0.84 GPa. Top right panel: oscillatory-part background-subtracted data above $H_{c2}$. Bottom left panel: FFT of the background subtracted data.  The lower frequency peak is a data analysis artifact coming from the polynomial background subtraction. }
		\label{pulsed.fig}	
\end{center}		
\end{figure}
Figure \ref{pulsed.fig} shows an example of data obtained in  a 70 T magnet  at 500 mK during a 150 ms pulse for a YBCO6.5 sample at 0.8 GPa.  The main panel shows the as-measured TDO frequency,  where the transition at about 25 T is identified as the superconducting critical field $H_{c2}$ in agreement with  other resistivity data in the literature \cite{Grissonnanche2014}.  Above this transition, clear QOs are observed in the data,  and thus we will use this as the marker for $H_{c2}$ throughout this work. The two traces correspond to the data taken during the up and down sweep in magnetic field,  with overlapping traces indicating that there is no heating of the sample during the pulse.  The background-subtracted data above $H_{c2}$ is shown in the top right-hand corner panel,  with one frequency clearly observed,  confirmed by the FFT (bottom left).  Similar data were obtained up to 7 GPa in a 85 T magnet,  shown in Sup.  Fig. \ref{SI2_Layout85T.fig}. \\

\subsubsection{Pressures greater than 18 GPa}
\begin{figure}[]
\begin{center}	
	\includegraphics[scale=0.29]{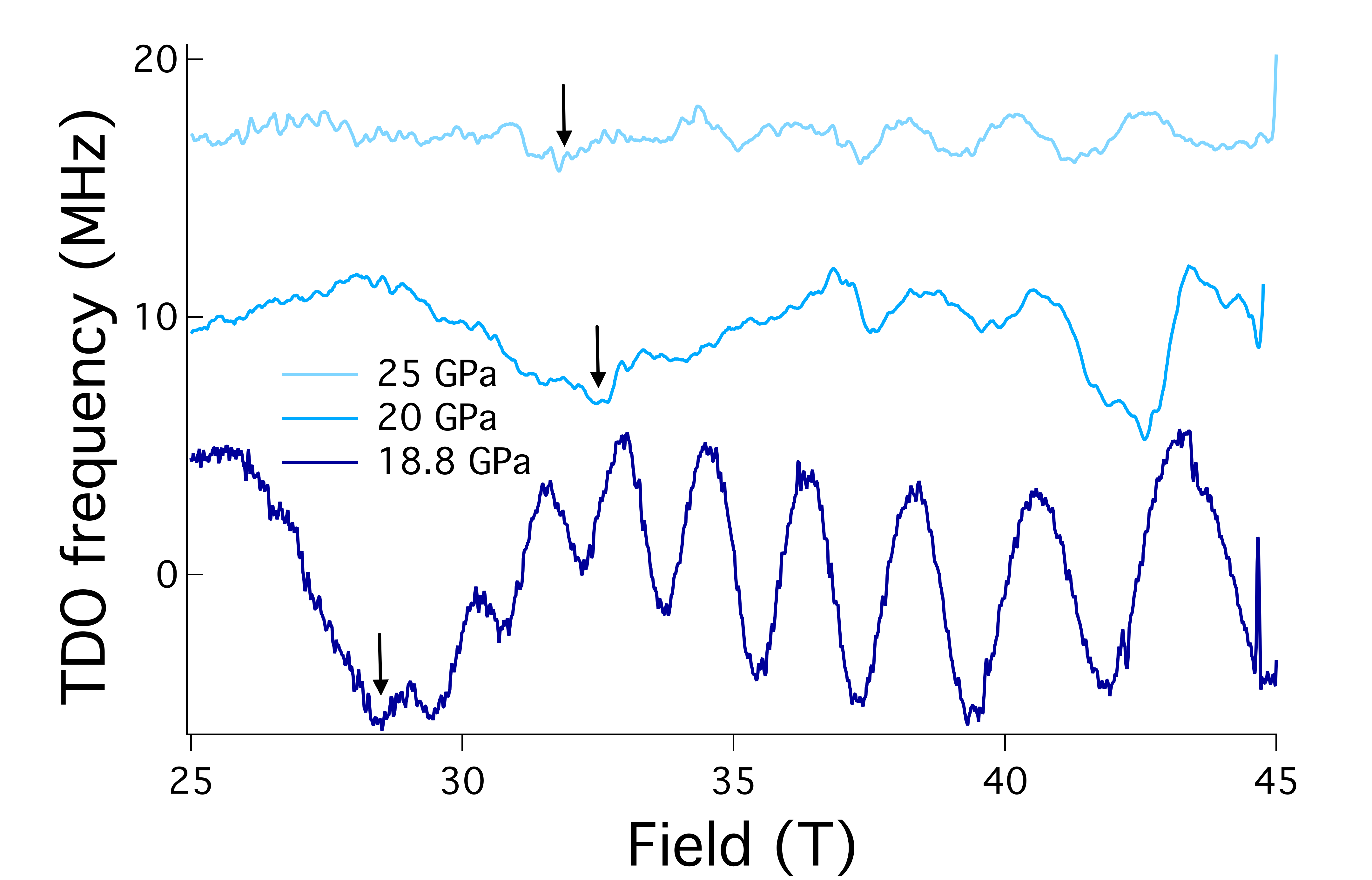}
		\caption{Base $^{3}$He temperature, background-subtracted TDO frequency as a function of field for three pressures.  The sample in the 20 GPa cell was bridged,  leading to poorer quality data.  $H_{c2}$ is indicated with arrows and determined by the onset field at which QOs appear. The traces are vertically shifted for clarity.}
		\label{fig4}	
\end{center}		
\end{figure}

\begin{figure}[]
\begin{center}	
	\includegraphics[scale=0.28]{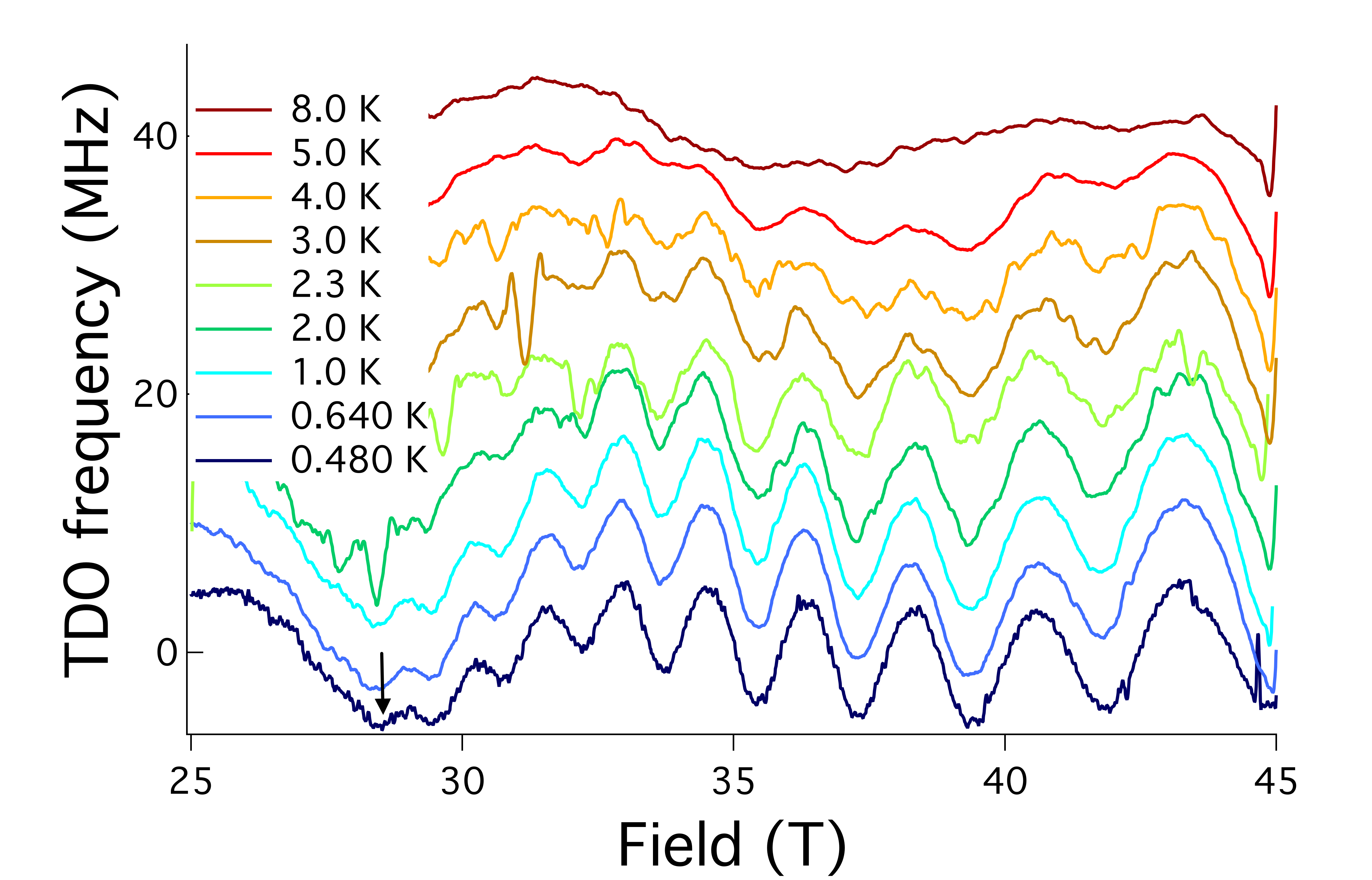}
		\caption{Evolution of the background-substracted TDO frequency as a function of field for various temperatures at 18.8 GPa at low temperature.  We assign the minimum in the signal before the onset of the QOs to $H_{c2}$, indicated by an arrow.  The various traces are vertically shifted for clarity. }
		\label{fig5}	
\end{center}		
\end{figure}
The estimated values of H$_{c2}$ as one approaches optimal doping are greater than the magnetic fields available in non-destructive pulsed magnets,  therefore higher pressures were used to drive the sample to the overdoped side.  For this purpose,  three different pressure cells loaded with a 4:1::methanol:ethanol mixture were taken to higher pressures (18.8,  20.0 and 25 GPa) at low temperatures.  In a related study,  Alireza $et$ $al. $ \cite{Alireza2017}  used a similar approach and reconstructed the overdoped side of the superconducting dome by applying pressures of about 25 GPa to an optimally doped sample $p$ = 0.17.  \\
Figure \ref{fig4} shows the background-subtracted TDO-frequency field dependence of our sample at base $^{3}$He temperatures. The 18.8 GPa cell shows a clear minimum in the signal at 28.4 T (indicated by an arrow) followed by the onset of QO.  Again, we identify this minimum as the critical field. This feature is not as pronounced for higher pressures, but QO are still observable. \\ Figure \ref{fig5} shows the evolution of the background-subtracted TDO frequency as a function of field for various temperatures for YBCO6.5 at 18.8 GPa.  The critical field is clearly suppressed with increasing temperature.  We plot $H_{c2}$ vs temperature for various pressures in Sup.  Fig.  \ref{SI5_Hc_T_dep.fig},  and show that the temperature dependence of the transition at 18.8 GPa follows the same trend.  \\

\subsection{High pressure fermiology}

\begin{figure}
\centering
\begin{subfigure}[]{}
\centering
\includegraphics[width=0.475\textwidth]{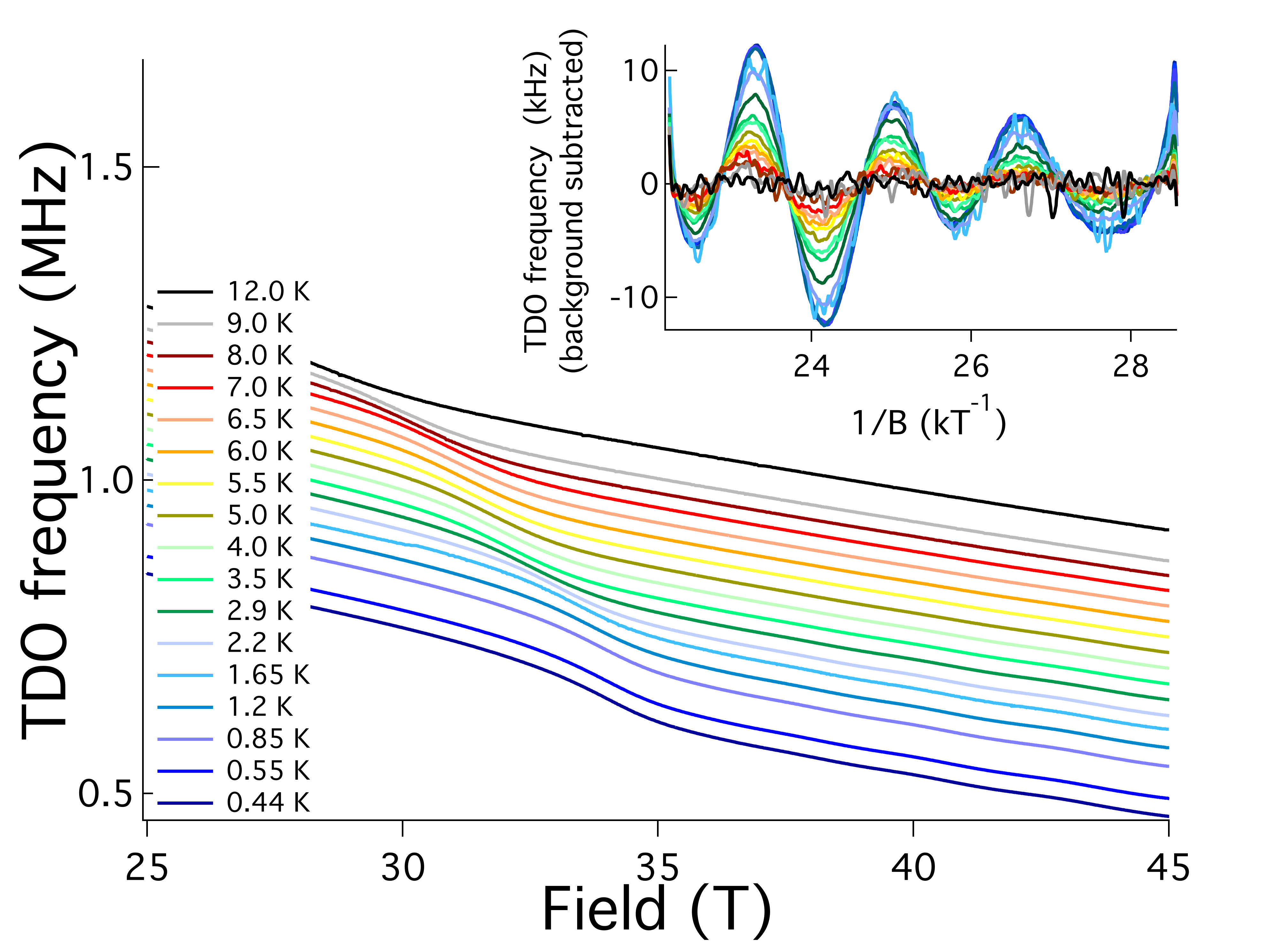} 
\end{subfigure}

\begin{subfigure}[]{}
\centering
\includegraphics[width=0.47\textwidth]{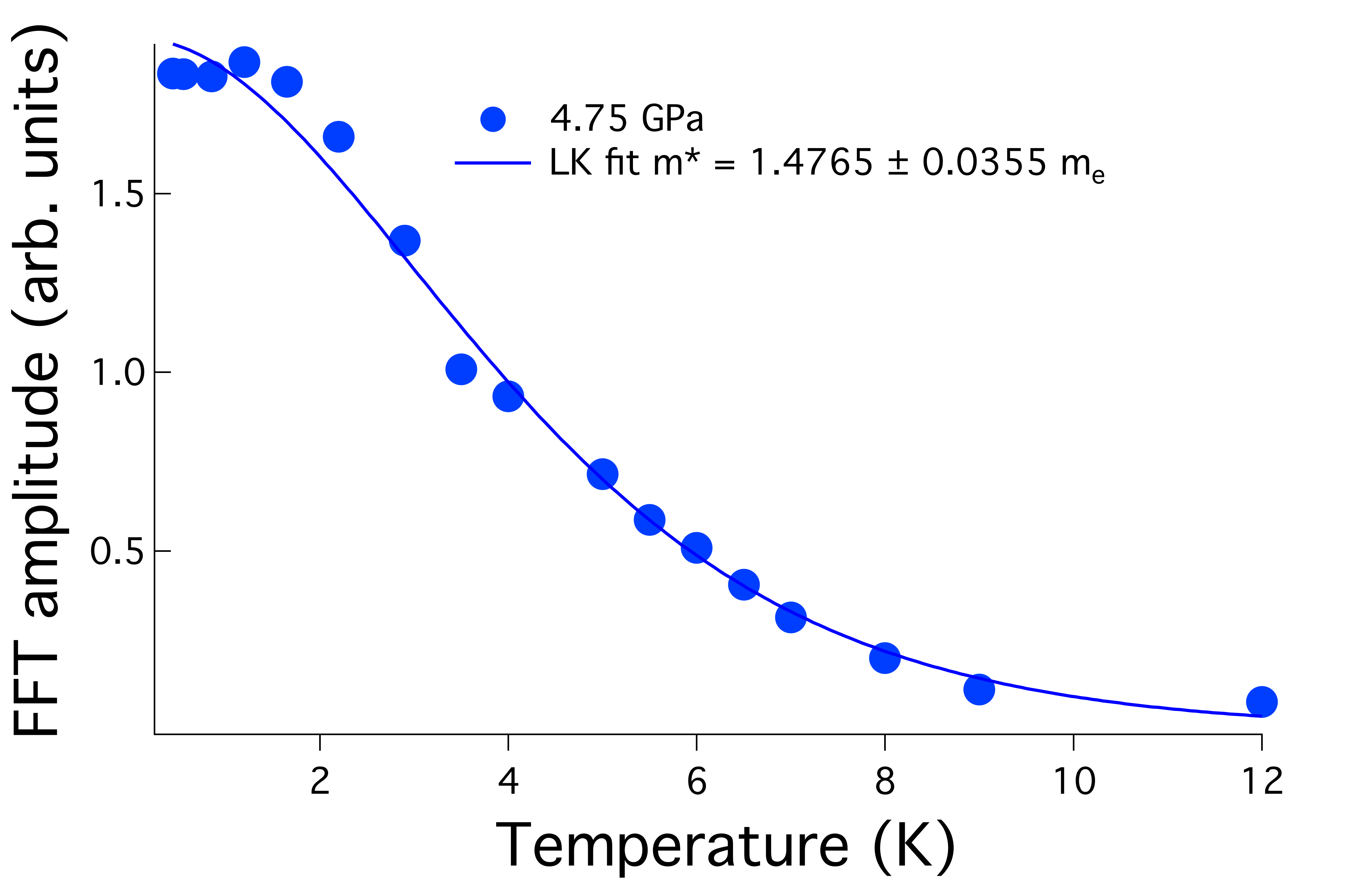} 
\end{subfigure}
 \caption{(a) TDO frequency at 4.8 GPa as a function of field for various temperatures, vertically shifted for clarity.  These traces are background-subtracted in the oscillatory region and plotted as a function of the inverse field in the insert.  The amplitude of the peak of the Fourier transform of these background-subtracted traces is fitted with a Lifshitz-Kosevich equation in (b),  yielding the effective mass of the carriers.  The deviation to the LK observed at temperatures below 1.4 K can be ascribed to heating due to the TDO.}
\label{fig6}
\end{figure}

For each pressure,  field sweeps were performed at various temperatures to study the temperature dependence of the QO amplitude, as shown in Fig.  \ref{fig6}(a) for 4.8 GPa and in Sup.  Fig. \ref{SI6.fig} for 25 GPa .  The oscillatory part is plotted as a function of 1/$B$, interpolated and smoothed with a Savitzky-Golay filter  \cite{Savitzky1964}. A smooth background is subtracted,  and the FFT of the result is taken using a Hamming window.  The amplitude A(T) of each Fourier transform peak is then plotted versus temperature as shown in Fig.  \ref{fig6}(b),  and fitted using the Lifshitz-Kosevitch (LK) theory (\ref{LKfit}) for the temperature damping factor \cite{Schoenberg2009}: 
\begin{align}
A(T) &= A_0\dfrac{\alpha m^{*}T/B}{\sinh (\alpha m^{*}T/B)}, \label{LKfit} 
\end{align}
where A$_0$ is a prefactor, $\alpha = 2\pi ^{2}k _{B}m_{e}/e\hbar = 14.69$ T/K,  $k _{B}$  and $\hbar$ the Boltzmann and Planck constants,  $e$ the electron charge,  and $B$ is taken as the harmonic mean of the oscillatory field range.  This LK fit gives the effective mass  $m^\star$ in units of the free-electron mass $m_e$ of the carrier for that orbit.\\
 
\begin{figure}[]
\begin{center}	
	\includegraphics[scale=0.29]{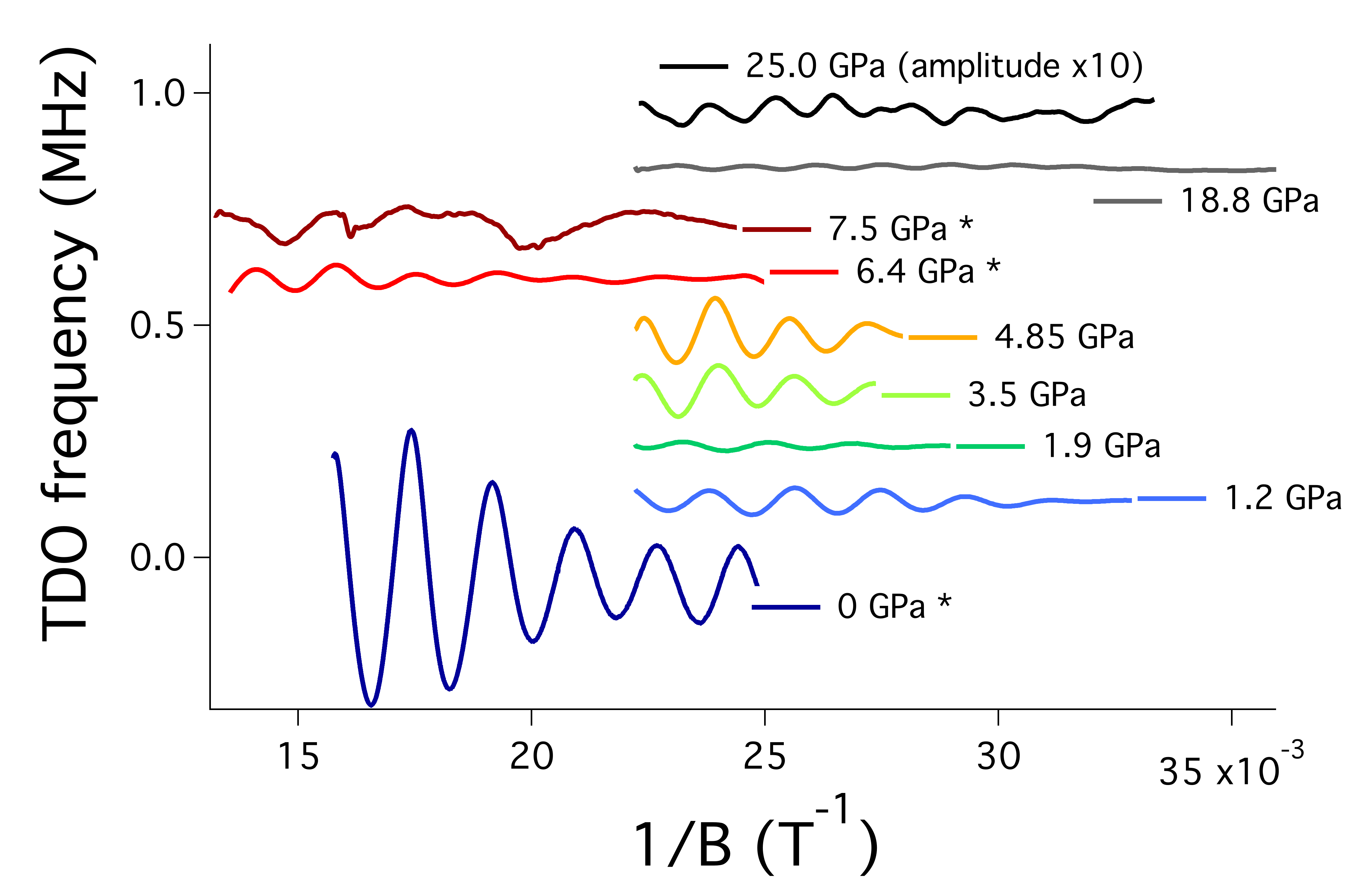}
		\caption{TDO frequency vs inverse field at base $^{3}$He temperatures for various pressures,  vertically shifted for clarity.  The data indicated with an asterisk were taken in pulsed magnetic fields.  }
		\label{fig7.fig}	
\end{center}		
\end{figure}

The observation of QO requires high-purity crystals with minimal scattering,  and hydrostatic conditions.  Despite non-hydrostatic conditions for the cells at pressure higher than 15 GPa, clear QO are observed in Fig.    \ref{fig4}, with their temperature evolution in Fig.  \ref{fig5}.  The high bulk modulus of YBCO makes it less susceptible to strains due to pressure gradients as suggested by Tateiwa \textit{et al. } \cite{Tateiwa2009}.  Figure \ref{fig7.fig} shows the oscillatory part of the TDO signal at about 400 mK for various pressures as a function of 1/$B$,  with the 25 GPa curve multiplied by 10 for better visibility.  Comparable SdH frequencies are observed throughout the pressure range.

\section{Discussion}
\begin{figure}
\centering
\begin{subfigure}[]{}
\centering
\includegraphics[width=0.45\textwidth]{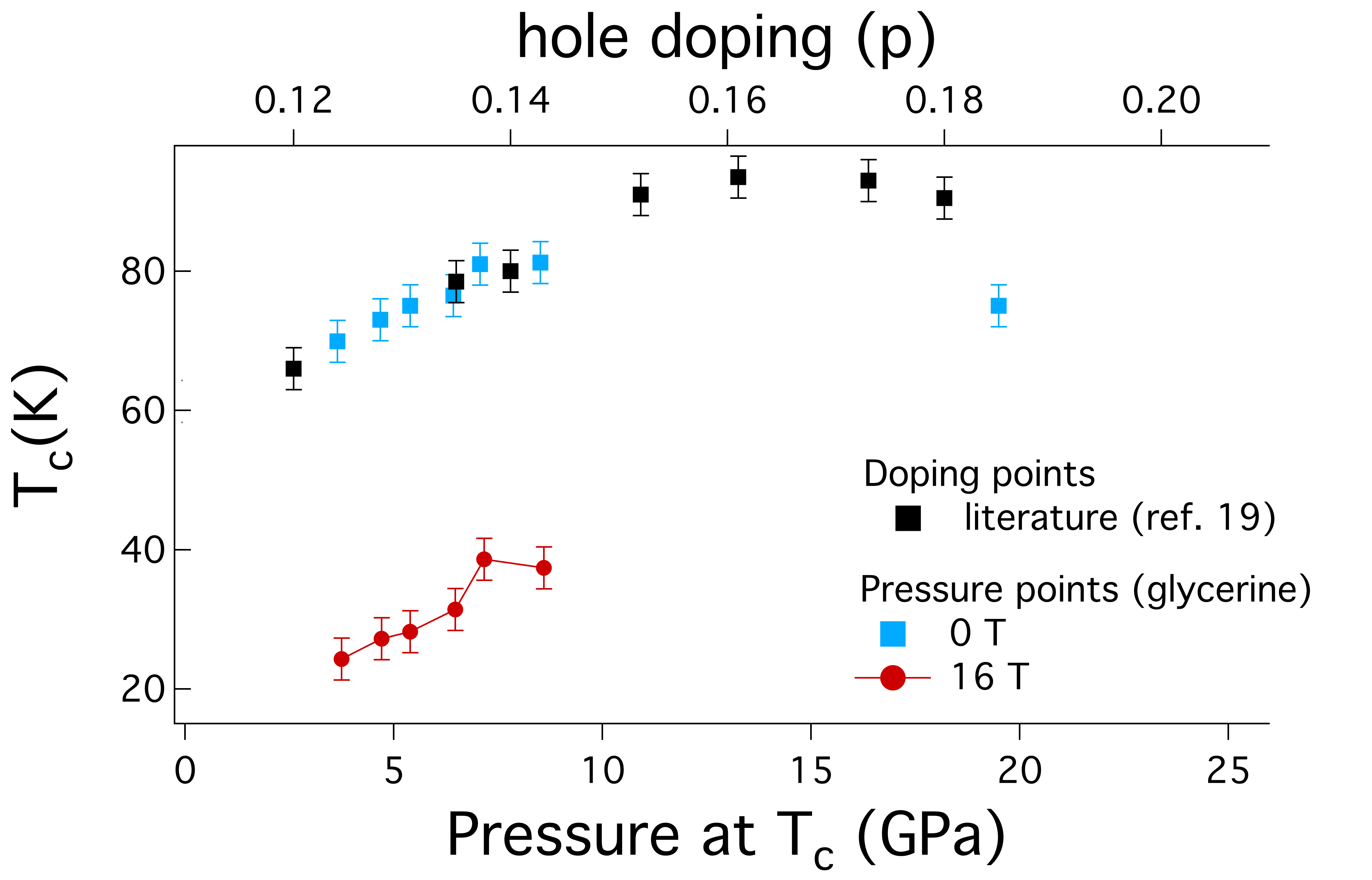} 
\end{subfigure}

\begin{subfigure}[]{}
\centering
\includegraphics[width=0.45\textwidth]{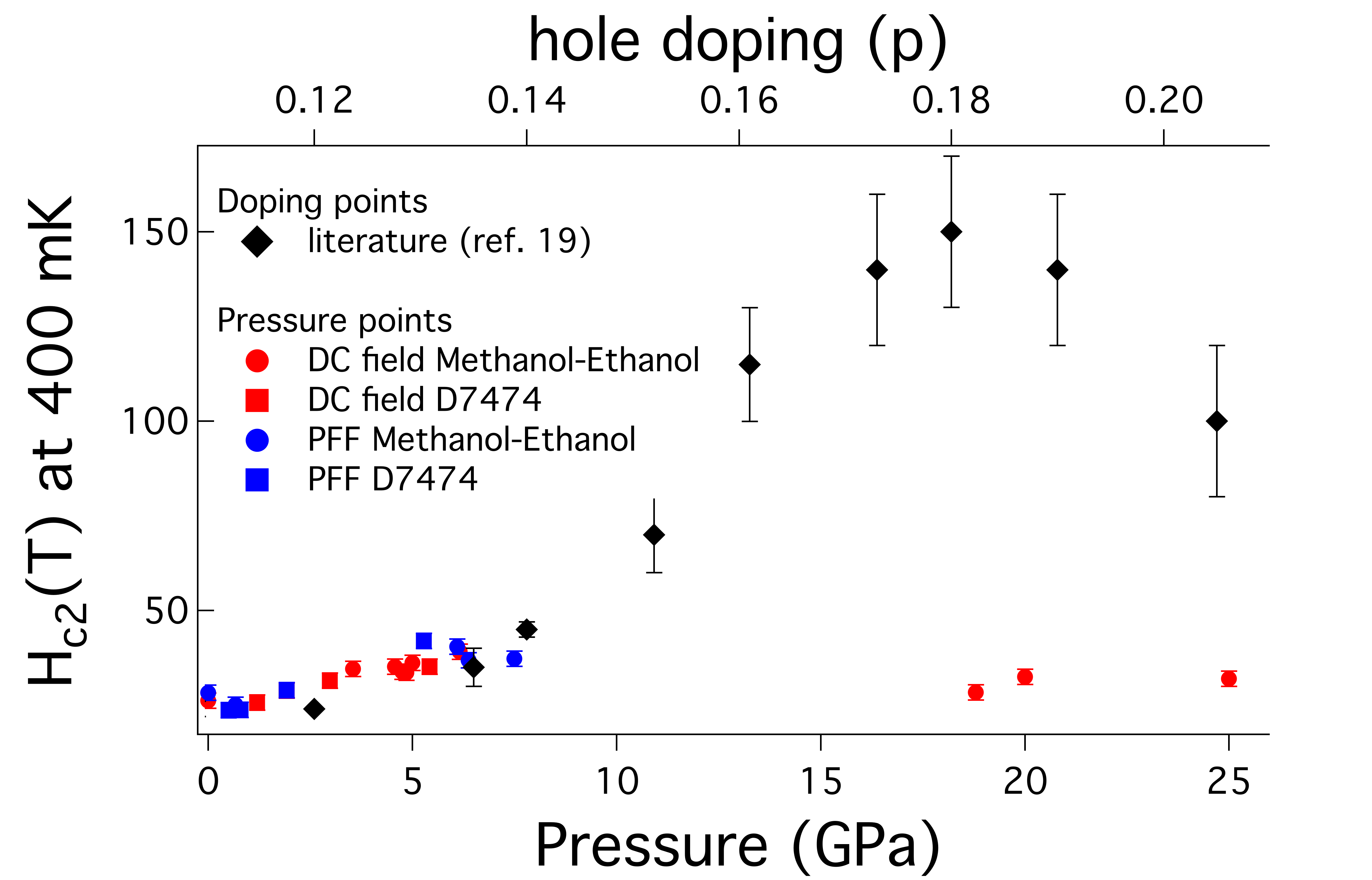} 
\end{subfigure}
 \caption{(a)$T_{c}$ as a function of pressure, at 0 and 16 T. The pressure is measured at $T$=$T_{c}$.  (b)$H_{c2}$ at base $^{3}$He temperature as a function of pressure.  The data points are separated depending on the pressure medium and magnet used.  $T_{c}$  and $H_{c2}$ vs doping data taken  from \cite{Grissonnanche2014}. }
\label{fig8.fig}
\end{figure}

Figure \ref{fig8.fig}(a) shows the evolution of $T_c$ with pressure,  increasing from 69.9 K at 3.65 GPa up to 81.2 K at 8.5 GPa.  This yields a $dT_c/dP$ $\approx$ 2.3 K/GPa,  in agreement with previous studies  (\cite{Sadewasser2000},  \cite{Cyr-Choiniere2018},  \cite{Jurkutat2023}).  
Previous attempts to explain the evolution of the superconducting transition with pressure P (\cite{Kawasaki2017}, \cite{Jurkutat2023}) invoke two main mechanisms described by:
\begin{equation}
\dfrac{dT_{c}}{dP} = (\dfrac{\partial T_{c} }{\partial n_{h}}) (\dfrac{\partial  n_{h} }{\partial P}) + (\dfrac{dT_{c}}{dP})_{intrinsic}, \label{Tcvsp} 
\end{equation}
The first term on the  right-hand side accounts for an increase of $T_c$  due to an increase in hole doping $n_{h}$ via a charge transfer between the CuO chains and the CuO plane. The second term encompasses intrinsic effects,  such as the increase of oxygen ordering in the CuO chains upon pressure. 
These intrinsic effects have been extensively studied \cite{Sadewasser2000}, and showed that the application of pressure at room temperature increases this ordering process, as opposed to applying the pressure at lower temperatures at which the oxygen chains are frozen.  Following the reasoning of Cyr-Choinière \textit{et al. } \cite{Cyr-Choiniere2018}, we compare our $dT_c/dP$ value to those reported by studies in which those effects are minimized, and find a  comparable  value of $dT_c/dP$.  In order to estimate the change of doping with pressure,  thus,  we neglect the intrinsic pressure effects on $T_c$. \\
Previous studies have attempted to quantify the change in doping with applied pressure: 
\begin{align}
n_{h}(P)&=n_{h,0} + \dfrac{dn_{h}(P)}{d\epsilon(P)}\epsilon(P), 
\end{align}
where $n_{h,0}$ is the initial hole doping (also denoted p),  and $\epsilon(P)$ is the change in unit-cell volume.
We apply the first-order Murnaghan equation of state \cite{Olsen1991},  \cite{Alireza2017}: 
\begin{align}
\dfrac{dn_{h}(P)}{d\epsilon(P)} &= n'_{h}(P)B_{0}(1 + \dfrac{B'_{0}}{B_{0}}P)^{1+1/B'_{0}}, \\ \label{Birch}
 \epsilon(p) &= 1-((1 + \dfrac{B'_{0}}{B_{0}}p)^{-1/B'_{0}}, 
\end{align}
where $B_{0}$ is the bulk modulus and $B^{'}_{0}$ its pressure derivative at zero pressure,  and $n^{'}_{h}$ = $dn_{h}(P)/d(P)$.  Comparing our point of $T_{c}$=81 K at 7.07 GPa,  with reported  $T_{c}$ values as a function of doping \cite{Grissonnanche2014},  we estimate the hole doping at that pressure to be of the order of $n_{h}$(7.07 GPa) = 0.142 holes/Cu.  Using approximation of values reported from x-ray diffraction data and calculations for various neighboring dopings of YBCO  \cite{Ludwig1992}, we use $B_{0}$ $\approx$ 140 $\pm$ 15 GPa,  and $B^{'}_{0}$ $\approx$ 5.5 $\pm$ 0.1,  which yields $dn_{h}(P)/d(P) (7.07 GPa)$ $\approx$ 0.37$\pm$ 0.1$\%$ hole/GPa,  in agreement with recent NMR studies \cite{Jurkutat2023}.  \\
Figure \ref{fig8.fig}(b) shows the evolution of $H_{c2}$ with pressure,  along with a comparison of the evolution in doping (data points from \cite{Grissonnanche2014} and references therein).  The pressure-doping comparison is consistent at pressures below 8 GPa.  Our initial decrease in $H_{c2}$ with a minimum at 0.8 GPa is consistent with the minimum in $H_{c2}$ observed near $p$=0.12.  The $H_{c2}$=28.4 T observed for 18.8 GPa places this sample on the overdoped side of the YBCO superconducting dome,  making it the first direct measurement of the critical field of overdoped YBCO.  The pressure-doping equivalence previously estimated  gives $p$ $\approx$0.18.  Our $H_{c2}$ value is,  however,  in discrepancy with previous measurements reporting $H_{c2}$=150 $\pm$ 20 T for this doping.  Comparing reported $H_{c2}$ values with ours would indicate a doping near $p$=0.24.  The pressure-doping equivalence calculated clearly breaks down at higher pressure,  possibly indicating that intrinsic effects, which we neglected, should be taken into account. Those effects can create strains along the different axes of the sample.  Several studies have looked into uniaxial compression effects on YBCO.  Kim \textit{et al.}, in particular \cite{Kim2018},  have suggested a three-dimensional charge order upon compression of the $a$-axis in zero magnetic field.  Thus,  one option to explain the discrepancy observed in $H_{c2}$ which could be investigated by high pressure NMR,  would be the existence of a competing charge order created at non-hydrostatic high pressures akin to strain.

\begin{figure}
\centering
\begin{subfigure}[]{}
\centering
\includegraphics[width=0.465\textwidth]{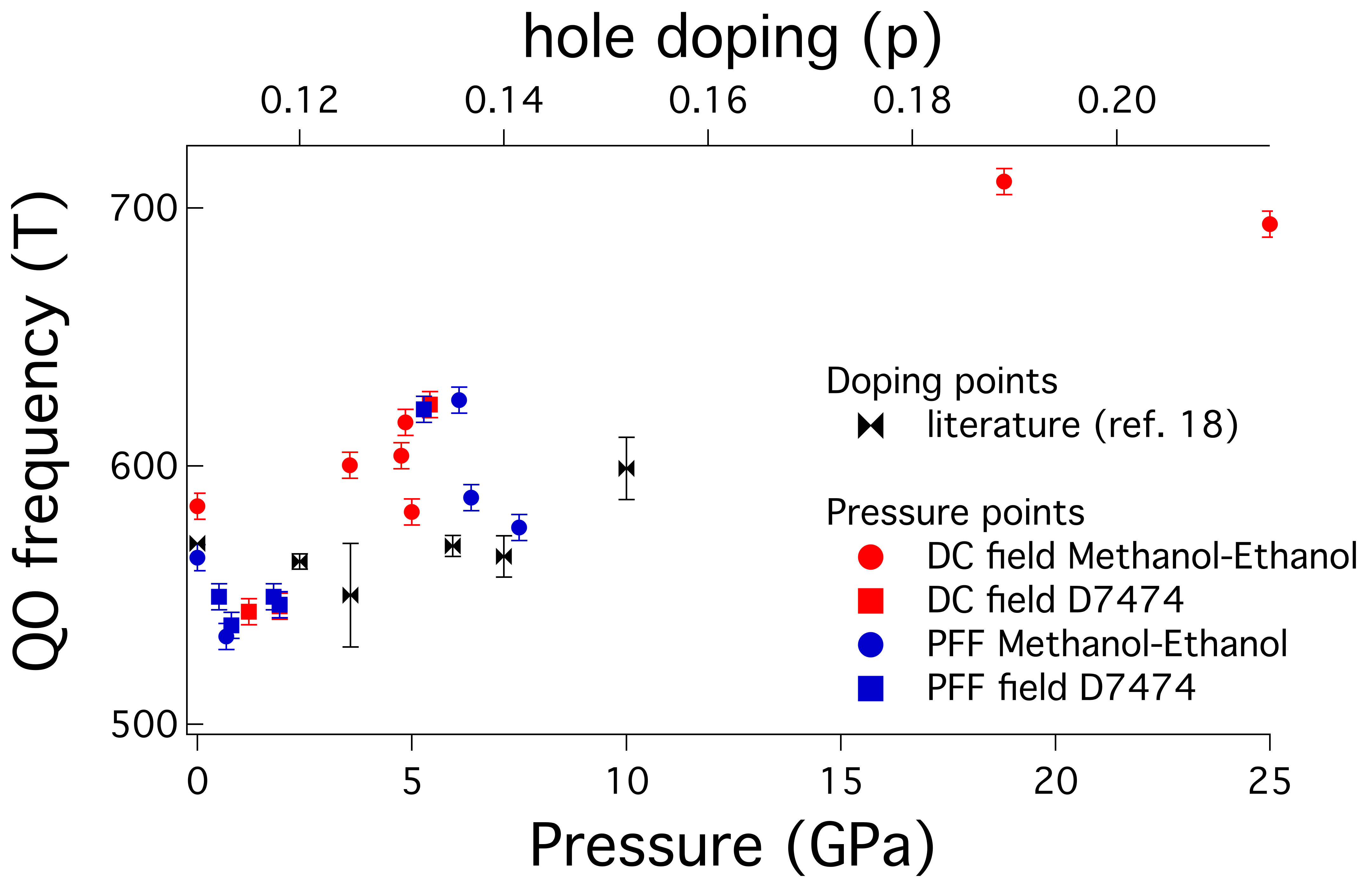} 
\end{subfigure}

\begin{subfigure}[]{}
\centering
\includegraphics[width=0.47\textwidth]{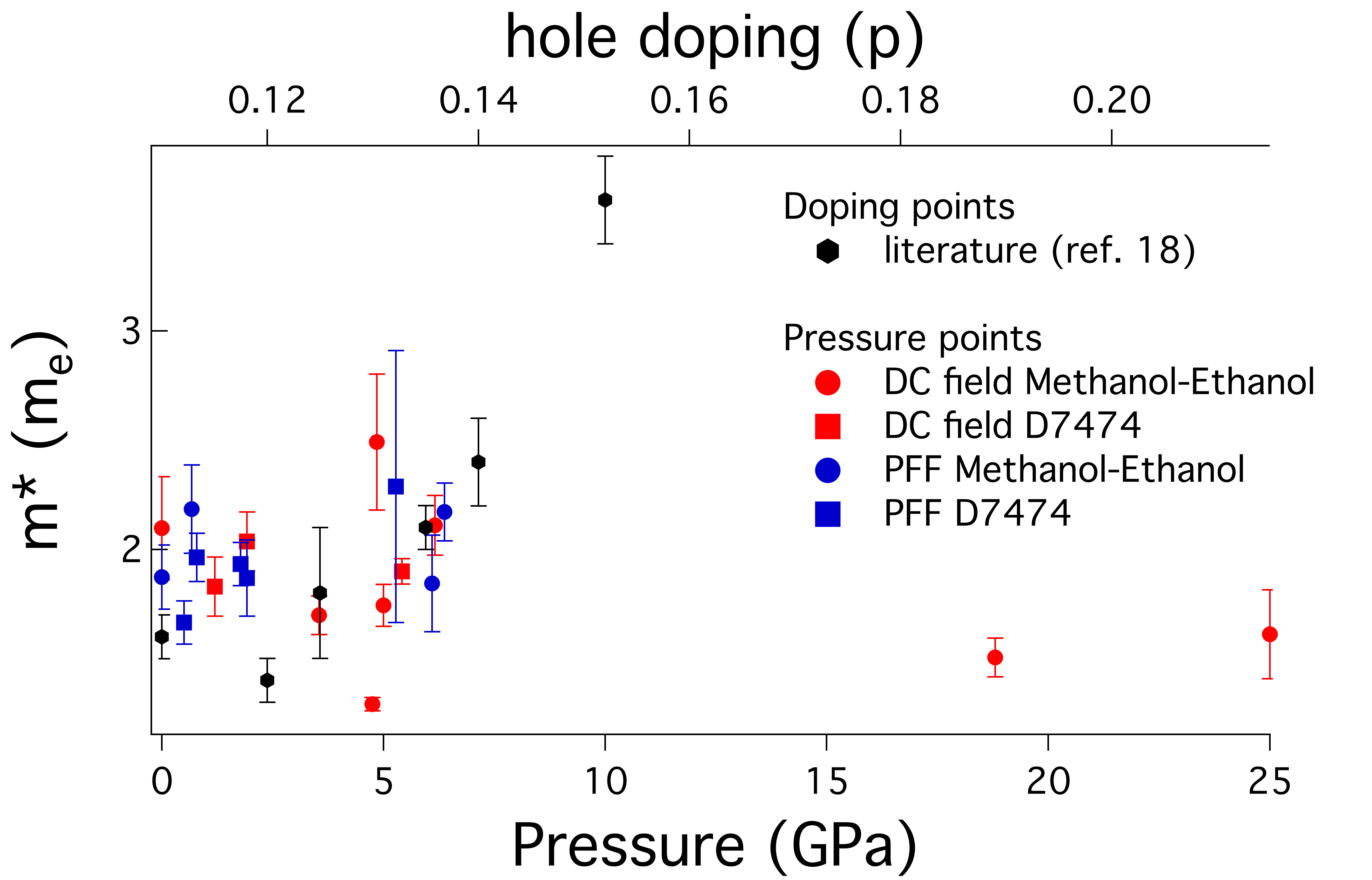} 
\end{subfigure}
 \caption{(a) QO frequencies measured vs pressure.  (b)Effective masses $m^{*}$  vs pressure. The doping data points are taken from \cite{Ramshaw317}.}
 \label{fig9}
\end{figure}

Figure \ref{fig9}(a) and (b) show, respectively, the evolution of the QO frequency and their effective masses with pressure. The data points for comparison with doping are taken from \cite{Ramshaw317} and references within.  The QO frequency range observed in our study,  550-650 T, is in excellent agreement with reported data. We observe a minimum in frequency at 0.8 GPa,  similar to that observed in $H_{c2}$, and a possible maximum near 6 GPa.  The values are consistent between DC field and pulsed-field measurements,  and no clear influence of the pressure medium is observed.  The $m^\star$ obtained from LK fits are in the range of 1.5-2.5$m_{e}$, also in agreement with reported data.  We note a minimum in $m^\star$ near 4 GPa.  Our low-pressure range does not extend far enough in $p$ to observe the divergence in $m^\star$ noted by Ramshaw \textit{et al.} \cite{Ramshaw317}.
The frequencies obtained for pressures greater than 18 GPa in the overdoped regime are,  however,  in complete discrepancy with the literature for the Tl analog.  We obtain 710 $\pm$ 10 T for 18.8 GPa,  and 694 $\pm$ 10 T at 25 GPa.  The corresponding $m^\star$ are 1.50 $\pm$ 0.08 $m_{e}$ and 1.6 $\pm$ 0.2 $m_{e}$,  respectively.  This is in disagreement with the proposed picture of a Fermi-surface reconstruction for overdoped samples based on measurements of Tl$_{2}$Ba$_{2}$CuO$_{6+\delta}$,  that were shown by angle-dependent magnetoresistance oscillations (AMRO) \cite{Hussey2003},  \cite{Abdel-Jawad2006}  and ARPES measurements   \cite{Damascelli2001},  \cite{Plate2005}  to have a large coherent three-dimensional Fermi surface.  Resistivity and torque measurements at ambient pressure \cite{Vignolle2011} have revealed QOs with a frequency of 18 kT $\pm$ 50 T and an effective mass $m^\star$ = 4.1 $\pm$ 1.0 m$_{e}$. The low $H_{c2}$ values at high pressure,  and negligible change in QOs frequency from low to high pressures  do not support the picture of a FSR across the superconducting dome,  and rather point to the existence of a charge order similar to the one observed at low pressure and on the underdoped side.

\section{Conclusion}
Our high pressure TDO measurements of YBCO6.5 have enabled direct measurement of $T_{c}$,  $H_{c2}$, and QOs, from which we derived QO frequencies and effective masses.  Measurements at pressures below 8 GPa show results in excellent agreement with previous doping studies, confirming the equivalence of pressure and doping.  We estimated the value for a charge transfer with pressure  and found it to be compatible with other data    \cite{Jurkutat2023}.  The pressure dependences of $H_{c2}$ and the QO frequency indicates also a suppression of the CO by applied pressure.  By applying pressures larger than 18 GPa,  the sample was driven to the overdoped regime on the other side of the superconducting dome,  as shown by the measured values of $T_{c}$ and $H_{c2}$.  However,  the QO frequency measured,  comparable to those obtained in the lower pressure range,  do not support the picture of a FSR in the overdoped regime of YBCO,  but rather indicate the existence of a charge order in competition with the superconducting state.  High pressure x-ray and NMR should shed light on the origin of this CO and inform on its links to the low pressure,  low-doping ones reported. 
\\
\\

\begin{acknowledgments}
This work was funded by the DOE/NNSA under DE-NA0001979 and was performed at the National High Magnetic Field Laboratory, which is supported by NSF Cooperative Agreement No. DMR-1157490 and by the State of Florida. Part of this work was supported as part of the Center for Actinide Science and Technology (CAST), an Energy Frontier Research Center funded by the U.S. Department of Energy, Office of Science, Basic Energy Sciences under Award No. DE-SC0016568,  and by the Deutsche Forschungsgemeinschaft (DFG) through the Würzburg-Dresden Cluster of Excellence on Complexity and Topology in Quantum Matter–ct.qmat (EXC 2147, Grant No. 390858490). We acknowledge the support of HLD at HZDR,  a member of the European Magnetic Field Laboratory (EMFL).  ADG,  WAC and SWT thank J. Wosnitza for financial support and careful review of the manuscript.  ADG acknowledges N. Doiron-Leyraud and S. Badoux for stimulating discussions, and C.  Agosta for crucial help in the data analysis.  SWT thanks D.  Schiferl for bringing the Parmax and Zylon to his attention. \end{acknowledgments}

\newpage
\newpage
\bibliographystyle{apsrev4-1}
\bibliography{YBCO_paper.bib}

\begin{thebibliography}{43}%
\makeatletter
\providecommand \@ifxundefined [1]{%
 \@ifx{#1\undefined}
}%
\providecommand \@ifnum [1]{%
 \ifnum #1\expandafter \@firstoftwo
 \else \expandafter \@secondoftwo
 \fi
}%
\providecommand \@ifx [1]{%
 \ifx #1\expandafter \@firstoftwo
 \else \expandafter \@secondoftwo
 \fi
}%
\providecommand \natexlab [1]{#1}%
\providecommand \enquote  [1]{``#1''}%
\providecommand \bibnamefont  [1]{#1}%
\providecommand \bibfnamefont [1]{#1}%
\providecommand \citenamefont [1]{#1}%
\providecommand \href@noop [0]{\@secondoftwo}%
\providecommand \href [0]{\begingroup \@sanitize@url \@href}%
\providecommand \@href[1]{\@@startlink{#1}\@@href}%
\providecommand \@@href[1]{\endgroup#1\@@endlink}%
\providecommand \@sanitize@url [0]{\catcode `\\12\catcode `\$12\catcode
  `\&12\catcode `\#12\catcode `\^12\catcode `\_12\catcode `\%12\relax}%
\providecommand \@@startlink[1]{}%
\providecommand \@@endlink[0]{}%
\providecommand \url  [0]{\begingroup\@sanitize@url \@url }%
\providecommand \@url [1]{\endgroup\@href {#1}{\urlprefix }}%
\providecommand \urlprefix  [0]{URL }%
\providecommand \Eprint [0]{\href }%
\providecommand \doibase [0]{http://dx.doi.org/}%
\providecommand \selectlanguage [0]{\@gobble}%
\providecommand \bibinfo  [0]{\@secondoftwo}%
\providecommand \bibfield  [0]{\@secondoftwo}%
\providecommand \translation [1]{[#1]}%
\providecommand \BibitemOpen [0]{}%
\providecommand \bibitemStop [0]{}%
\providecommand \bibitemNoStop [0]{.\EOS\space}%
\providecommand \EOS [0]{\spacefactor3000\relax}%
\providecommand \BibitemShut  [1]{\csname bibitem#1\endcsname}%
\let\auto@bib@innerbib\@empty
\bibitem [{\citenamefont {Wu}\ \emph {et~al.}(1987)\citenamefont {Wu},
  \citenamefont {Ashburn}, \citenamefont {Torng}, \citenamefont {Hor},
  \citenamefont {Meng}, \citenamefont {Gao}, \citenamefont {Huang},
  \citenamefont {Wang},\ and\ \citenamefont {Chu}}]{Wu1987}%
  \BibitemOpen
  \bibfield  {author} {\bibinfo {author} {\bibfnamefont {M.~K.}\ \bibnamefont
  {Wu}}, \bibinfo {author} {\bibfnamefont {J.~R.}\ \bibnamefont {Ashburn}},
  \bibinfo {author} {\bibfnamefont {C.~J.}\ \bibnamefont {Torng}}, \bibinfo
  {author} {\bibfnamefont {P.~H.}\ \bibnamefont {Hor}}, \bibinfo {author}
  {\bibfnamefont {R.~L.}\ \bibnamefont {Meng}}, \bibinfo {author}
  {\bibfnamefont {L.}~\bibnamefont {Gao}}, \bibinfo {author} {\bibfnamefont
  {Z.~J.}\ \bibnamefont {Huang}}, \bibinfo {author} {\bibfnamefont {Y.~Q.}\
  \bibnamefont {Wang}}, \ and\ \bibinfo {author} {\bibfnamefont {C.~W.}\
  \bibnamefont {Chu}},\ }\href
  {https://link.aps.org/doi/10.1103/PhysRevLett.58.908} {\bibfield  {journal}
  {\bibinfo  {journal} {Phys. Rev. Lett.}\ }\textbf {\bibinfo {volume} {58}},\
  \bibinfo {pages} {908} (\bibinfo {year} {1987})}\BibitemShut {NoStop}%
\bibitem [{\citenamefont {Taillefer}(2009)}]{Taillefer2009}%
  \BibitemOpen
  \bibfield  {author} {\bibinfo {author} {\bibfnamefont {L.}~\bibnamefont
  {Taillefer}},\ }\href {http://stacks.iop.org/0953-8984/21/i=16/a=164212}
  {\bibfield  {journal} {\bibinfo  {journal} {J. Condens. Matter Phys.}\
  }\textbf {\bibinfo {volume} {21}},\ \bibinfo {pages} {164212} (\bibinfo
  {year} {2009})}\BibitemShut {NoStop}%
\bibitem [{\citenamefont {LeBoeuf}\ \emph {et~al.}(2007)\citenamefont
  {LeBoeuf}, \citenamefont {Doiron-Leyraud}, \citenamefont {Levallois},
  \citenamefont {Daou}, \citenamefont {Bonnemaison}, \citenamefont {Hussey},
  \citenamefont {Balicas}, \citenamefont {Ramshaw}, \citenamefont {Liang},
  \citenamefont {Bonn}, \citenamefont {Hardy}, \citenamefont {Adachi},
  \citenamefont {Proust},\ and\ \citenamefont {Taillefer}}]{LeBoeuf2007}%
  \BibitemOpen
  \bibfield  {author} {\bibinfo {author} {\bibfnamefont {D.}~\bibnamefont
  {LeBoeuf}}, \bibinfo {author} {\bibfnamefont {N.}~\bibnamefont
  {Doiron-Leyraud}}, \bibinfo {author} {\bibfnamefont {J.}~\bibnamefont
  {Levallois}}, \bibinfo {author} {\bibfnamefont {R.}~\bibnamefont {Daou}},
  \bibinfo {author} {\bibfnamefont {J.-B.}\ \bibnamefont {Bonnemaison}},
  \bibinfo {author} {\bibfnamefont {N.~E.}\ \bibnamefont {Hussey}}, \bibinfo
  {author} {\bibfnamefont {L.}~\bibnamefont {Balicas}}, \bibinfo {author}
  {\bibfnamefont {B.~J.}\ \bibnamefont {Ramshaw}}, \bibinfo {author}
  {\bibfnamefont {R.}~\bibnamefont {Liang}}, \bibinfo {author} {\bibfnamefont
  {D.~A.}\ \bibnamefont {Bonn}}, \bibinfo {author} {\bibfnamefont {W.~N.}\
  \bibnamefont {Hardy}}, \bibinfo {author} {\bibfnamefont {S.}~\bibnamefont
  {Adachi}}, \bibinfo {author} {\bibfnamefont {C.}~\bibnamefont {Proust}}, \
  and\ \bibinfo {author} {\bibfnamefont {L.}~\bibnamefont {Taillefer}},\ }\href
  {http://dx.doi.org/10.1038/nature06332} {\bibfield  {journal} {\bibinfo
  {journal} {Nature}\ }\textbf {\bibinfo {volume} {450}},\ \bibinfo {pages}
  {533} (\bibinfo {year} {2007})}\BibitemShut {NoStop}%
\bibitem [{\citenamefont {LeBoeuf}\ \emph {et~al.}(2011)\citenamefont
  {LeBoeuf}, \citenamefont {Doiron-Leyraud}, \citenamefont {Vignolle},
  \citenamefont {Sutherland}, \citenamefont {Ramshaw}, \citenamefont
  {Levallois}, \citenamefont {Daou}, \citenamefont {Lalibert{\'e}},
  \citenamefont {Cyr-Choini{\`e}re}, \citenamefont {Chang}, \citenamefont {Jo},
  \citenamefont {Balicas}, \citenamefont {Liang}, \citenamefont {Bonn},
  \citenamefont {Hardy}, \citenamefont {Proust},\ and\ \citenamefont
  {Taillefer}}]{LeBoeuf2011}%
  \BibitemOpen
  \bibfield  {author} {\bibinfo {author} {\bibfnamefont {D.}~\bibnamefont
  {LeBoeuf}}, \bibinfo {author} {\bibfnamefont {N.}~\bibnamefont
  {Doiron-Leyraud}}, \bibinfo {author} {\bibfnamefont {B.}~\bibnamefont
  {Vignolle}}, \bibinfo {author} {\bibfnamefont {M.}~\bibnamefont
  {Sutherland}}, \bibinfo {author} {\bibfnamefont {B.~J.}\ \bibnamefont
  {Ramshaw}}, \bibinfo {author} {\bibfnamefont {J.}~\bibnamefont {Levallois}},
  \bibinfo {author} {\bibfnamefont {R.}~\bibnamefont {Daou}}, \bibinfo {author}
  {\bibfnamefont {F.}~\bibnamefont {Lalibert{\'e}}}, \bibinfo {author}
  {\bibfnamefont {O.}~\bibnamefont {Cyr-Choini{\`e}re}}, \bibinfo {author}
  {\bibfnamefont {J.}~\bibnamefont {Chang}}, \bibinfo {author} {\bibfnamefont
  {Y.~J.}\ \bibnamefont {Jo}}, \bibinfo {author} {\bibfnamefont
  {L.}~\bibnamefont {Balicas}}, \bibinfo {author} {\bibfnamefont
  {R.}~\bibnamefont {Liang}}, \bibinfo {author} {\bibfnamefont {D.~A.}\
  \bibnamefont {Bonn}}, \bibinfo {author} {\bibfnamefont {W.~N.}\ \bibnamefont
  {Hardy}}, \bibinfo {author} {\bibfnamefont {C.}~\bibnamefont {Proust}}, \
  and\ \bibinfo {author} {\bibfnamefont {L.}~\bibnamefont {Taillefer}},\ }\href
  {https://link.aps.org/doi/10.1103/PhysRevB.83.054506} {\bibfield  {journal}
  {\bibinfo  {journal} {Phys. Rev. B}\ }\textbf {\bibinfo {volume} {83}},\
  \bibinfo {pages} {054506} (\bibinfo {year} {2011})}\BibitemShut {NoStop}%
\bibitem [{\citenamefont {Chang}\ \emph {et~al.}(2010)\citenamefont {Chang},
  \citenamefont {Daou}, \citenamefont {Proust}, \citenamefont {LeBoeuf},
  \citenamefont {Doiron-Leyraud}, \citenamefont {Laliberté}, \citenamefont
  {Pingault}, \citenamefont {Ramshaw}, \citenamefont {Liang}, \citenamefont
  {Bonn}, \citenamefont {Hardy}, \citenamefont {Takagi}, \citenamefont
  {Antunes}, \citenamefont {Sheikin}, \citenamefont {Behnia},\ and\
  \citenamefont {Taillefer}}]{Chang2010}%
  \BibitemOpen
  \bibfield  {author} {\bibinfo {author} {\bibfnamefont {J.}~\bibnamefont
  {Chang}}, \bibinfo {author} {\bibfnamefont {R.}~\bibnamefont {Daou}},
  \bibinfo {author} {\bibfnamefont {C.}~\bibnamefont {Proust}}, \bibinfo
  {author} {\bibfnamefont {D.}~\bibnamefont {LeBoeuf}}, \bibinfo {author}
  {\bibfnamefont {N.}~\bibnamefont {Doiron-Leyraud}}, \bibinfo {author}
  {\bibfnamefont {F.}~\bibnamefont {Laliberté}}, \bibinfo {author}
  {\bibfnamefont {B.}~\bibnamefont {Pingault}}, \bibinfo {author}
  {\bibfnamefont {B.~J.}\ \bibnamefont {Ramshaw}}, \bibinfo {author}
  {\bibfnamefont {R.}~\bibnamefont {Liang}}, \bibinfo {author} {\bibfnamefont
  {D.~A.}\ \bibnamefont {Bonn}}, \bibinfo {author} {\bibfnamefont {W.~N.}\
  \bibnamefont {Hardy}}, \bibinfo {author} {\bibfnamefont {H.}~\bibnamefont
  {Takagi}}, \bibinfo {author} {\bibfnamefont {A.~B.}\ \bibnamefont {Antunes}},
  \bibinfo {author} {\bibfnamefont {I.}~\bibnamefont {Sheikin}}, \bibinfo
  {author} {\bibfnamefont {K.}~\bibnamefont {Behnia}}, \ and\ \bibinfo {author}
  {\bibfnamefont {L.}~\bibnamefont {Taillefer}},\ }\href
  {https://link.aps.org/doi/10.1103/PhysRevLett.104.057005} {\bibfield
  {journal} {\bibinfo  {journal} {Phys. Rev. Lett.}\ }\textbf {\bibinfo
  {volume} {104}},\ \bibinfo {pages} {057005} (\bibinfo {year}
  {2010})}\BibitemShut {NoStop}%
\bibitem [{\citenamefont {Wu}\ \emph {et~al.}(2011)\citenamefont {Wu},
  \citenamefont {Mayaffre}, \citenamefont {Kramer}, \citenamefont
  {Horvati{\'c}}, \citenamefont {Berthier}, \citenamefont {Hardy},
  \citenamefont {Liang}, \citenamefont {Bonn},\ and\ \citenamefont
  {Julien}}]{Wu2011}%
  \BibitemOpen
  \bibfield  {author} {\bibinfo {author} {\bibfnamefont {T.}~\bibnamefont
  {Wu}}, \bibinfo {author} {\bibfnamefont {H.}~\bibnamefont {Mayaffre}},
  \bibinfo {author} {\bibfnamefont {S.}~\bibnamefont {Kramer}}, \bibinfo
  {author} {\bibfnamefont {M.}~\bibnamefont {Horvati{\'c}}}, \bibinfo {author}
  {\bibfnamefont {C.}~\bibnamefont {Berthier}}, \bibinfo {author}
  {\bibfnamefont {W.~N.}\ \bibnamefont {Hardy}}, \bibinfo {author}
  {\bibfnamefont {R.}~\bibnamefont {Liang}}, \bibinfo {author} {\bibfnamefont
  {D.~A.}\ \bibnamefont {Bonn}}, \ and\ \bibinfo {author} {\bibfnamefont
  {M.-H.}\ \bibnamefont {Julien}},\ }\href
  {http://dx.doi.org/10.1038/nature10345} {\bibfield  {journal} {\bibinfo
  {journal} {Nature}\ }\textbf {\bibinfo {volume} {477}},\ \bibinfo {pages}
  {191} (\bibinfo {year} {2011})}\BibitemShut {NoStop}%
\bibitem [{\citenamefont {Chang}\ \emph {et~al.}(2012)\citenamefont {Chang},
  \citenamefont {Blackburn}, \citenamefont {Holmes}, \citenamefont
  {Christensen}, \citenamefont {Larsen}, \citenamefont {Mesot}, \citenamefont
  {Liang}, \citenamefont {Bonn}, \citenamefont {Hardy}, \citenamefont
  {Watenphul}, \citenamefont {Zimmermann}, \citenamefont {Forgan},\ and\
  \citenamefont {Hayden}}]{Chang2012}%
  \BibitemOpen
  \bibfield  {author} {\bibinfo {author} {\bibfnamefont {J.}~\bibnamefont
  {Chang}}, \bibinfo {author} {\bibfnamefont {E.}~\bibnamefont {Blackburn}},
  \bibinfo {author} {\bibfnamefont {A.~T.}\ \bibnamefont {Holmes}}, \bibinfo
  {author} {\bibfnamefont {N.~B.}\ \bibnamefont {Christensen}}, \bibinfo
  {author} {\bibfnamefont {J.}~\bibnamefont {Larsen}}, \bibinfo {author}
  {\bibfnamefont {J.}~\bibnamefont {Mesot}}, \bibinfo {author} {\bibfnamefont
  {R.}~\bibnamefont {Liang}}, \bibinfo {author} {\bibfnamefont {D.~A.}\
  \bibnamefont {Bonn}}, \bibinfo {author} {\bibfnamefont {W.~N.}\ \bibnamefont
  {Hardy}}, \bibinfo {author} {\bibfnamefont {A.}~\bibnamefont {Watenphul}},
  \bibinfo {author} {\bibfnamefont {M.~v.}\ \bibnamefont {Zimmermann}},
  \bibinfo {author} {\bibfnamefont {E.~M.}\ \bibnamefont {Forgan}}, \ and\
  \bibinfo {author} {\bibfnamefont {S.~M.}\ \bibnamefont {Hayden}},\ }\href
  {https://doi.org/10.1038/nphys2456} {\bibfield  {journal} {\bibinfo
  {journal} {Nat. Phys.}\ }\textbf {\bibinfo {volume} {8}},\ \bibinfo {pages}
  {871} (\bibinfo {year} {2012})}\BibitemShut {NoStop}%
\bibitem [{\citenamefont {Gerber}\ \emph {et~al.}(2015)\citenamefont {Gerber},
  \citenamefont {Jang}, \citenamefont {Nojiri}, \citenamefont {Matsuzawa},
  \citenamefont {Yasumura}, \citenamefont {Bonn}, \citenamefont {Liang},
  \citenamefont {Hardy}, \citenamefont {Islam}, \citenamefont {Mehta},
  \citenamefont {Song}, \citenamefont {Sikorski}, \citenamefont {Stefanescu},
  \citenamefont {Feng}, \citenamefont {Kivelson}, \citenamefont {Devereaux},
  \citenamefont {Shen}, \citenamefont {Kao}, \citenamefont {Lee}, \citenamefont
  {Zhu},\ and\ \citenamefont {Lee}}]{Gerber2015}%
  \BibitemOpen
  \bibfield  {author} {\bibinfo {author} {\bibfnamefont {S.}~\bibnamefont
  {Gerber}}, \bibinfo {author} {\bibfnamefont {H.}~\bibnamefont {Jang}},
  \bibinfo {author} {\bibfnamefont {H.}~\bibnamefont {Nojiri}}, \bibinfo
  {author} {\bibfnamefont {S.}~\bibnamefont {Matsuzawa}}, \bibinfo {author}
  {\bibfnamefont {H.}~\bibnamefont {Yasumura}}, \bibinfo {author}
  {\bibfnamefont {D.~A.}\ \bibnamefont {Bonn}}, \bibinfo {author}
  {\bibfnamefont {R.}~\bibnamefont {Liang}}, \bibinfo {author} {\bibfnamefont
  {W.~N.}\ \bibnamefont {Hardy}}, \bibinfo {author} {\bibfnamefont
  {Z.}~\bibnamefont {Islam}}, \bibinfo {author} {\bibfnamefont
  {A.}~\bibnamefont {Mehta}}, \bibinfo {author} {\bibfnamefont
  {S.}~\bibnamefont {Song}}, \bibinfo {author} {\bibfnamefont {M.}~\bibnamefont
  {Sikorski}}, \bibinfo {author} {\bibfnamefont {D.}~\bibnamefont
  {Stefanescu}}, \bibinfo {author} {\bibfnamefont {Y.}~\bibnamefont {Feng}},
  \bibinfo {author} {\bibfnamefont {S.~A.}\ \bibnamefont {Kivelson}}, \bibinfo
  {author} {\bibfnamefont {T.~P.}\ \bibnamefont {Devereaux}}, \bibinfo {author}
  {\bibfnamefont {Z.-X.}\ \bibnamefont {Shen}}, \bibinfo {author}
  {\bibfnamefont {C.-C.}\ \bibnamefont {Kao}}, \bibinfo {author} {\bibfnamefont
  {W.-S.}\ \bibnamefont {Lee}}, \bibinfo {author} {\bibfnamefont
  {D.}~\bibnamefont {Zhu}}, \ and\ \bibinfo {author} {\bibfnamefont {J.-S.}\
  \bibnamefont {Lee}},\ }\href
  {http://science.sciencemag.org/content/350/6263/949.abstract} {\bibfield
  {journal} {\bibinfo  {journal} {Science}\ }\textbf {\bibinfo {volume}
  {350}},\ \bibinfo {pages} {949} (\bibinfo {year} {2015})}\BibitemShut
  {NoStop}%
\bibitem [{\citenamefont {Lalibert{\'e}}\ \emph {et~al.}(2018)\citenamefont
  {Lalibert{\'e}}, \citenamefont {Frachet}, \citenamefont {Benhabib},
  \citenamefont {Borgnic}, \citenamefont {Loew}, \citenamefont {Porras},
  \citenamefont {Le~Tacon}, \citenamefont {Keimer}, \citenamefont {Wiedmann},
  \citenamefont {Proust},\ and\ \citenamefont {LeBoeuf}}]{Laliberte2018}%
  \BibitemOpen
  \bibfield  {author} {\bibinfo {author} {\bibfnamefont {F.}~\bibnamefont
  {Lalibert{\'e}}}, \bibinfo {author} {\bibfnamefont {M.}~\bibnamefont
  {Frachet}}, \bibinfo {author} {\bibfnamefont {S.}~\bibnamefont {Benhabib}},
  \bibinfo {author} {\bibfnamefont {B.}~\bibnamefont {Borgnic}}, \bibinfo
  {author} {\bibfnamefont {T.}~\bibnamefont {Loew}}, \bibinfo {author}
  {\bibfnamefont {J.}~\bibnamefont {Porras}}, \bibinfo {author} {\bibfnamefont
  {M.}~\bibnamefont {Le~Tacon}}, \bibinfo {author} {\bibfnamefont
  {B.}~\bibnamefont {Keimer}}, \bibinfo {author} {\bibfnamefont
  {S.}~\bibnamefont {Wiedmann}}, \bibinfo {author} {\bibfnamefont
  {C.}~\bibnamefont {Proust}}, \ and\ \bibinfo {author} {\bibfnamefont
  {D.}~\bibnamefont {LeBoeuf}},\ }\href
  {https://doi.org/10.1038/s41535-018-0084-5} {\bibfield  {journal} {\bibinfo
  {journal} {npj Quantum Mater.}\ }\textbf {\bibinfo {volume} {3}},\ \bibinfo
  {pages} {11} (\bibinfo {year} {2018})}\BibitemShut {NoStop}%
\bibitem [{\citenamefont {Wu}\ \emph {et~al.}(2013)\citenamefont {Wu},
  \citenamefont {Mayaffre}, \citenamefont {Kr{\"a}mer}, \citenamefont
  {Horvati{\'c}}, \citenamefont {Berthier}, \citenamefont {Kuhns},
  \citenamefont {Reyes}, \citenamefont {Liang}, \citenamefont {Hardy},
  \citenamefont {Bonn},\ and\ \citenamefont {Julien}}]{Wu2013}%
  \BibitemOpen
  \bibfield  {author} {\bibinfo {author} {\bibfnamefont {T.}~\bibnamefont
  {Wu}}, \bibinfo {author} {\bibfnamefont {H.}~\bibnamefont {Mayaffre}},
  \bibinfo {author} {\bibfnamefont {S.}~\bibnamefont {Kr{\"a}mer}}, \bibinfo
  {author} {\bibfnamefont {M.}~\bibnamefont {Horvati{\'c}}}, \bibinfo {author}
  {\bibfnamefont {C.}~\bibnamefont {Berthier}}, \bibinfo {author}
  {\bibfnamefont {P.~L.}\ \bibnamefont {Kuhns}}, \bibinfo {author}
  {\bibfnamefont {A.~P.}\ \bibnamefont {Reyes}}, \bibinfo {author}
  {\bibfnamefont {R.}~\bibnamefont {Liang}}, \bibinfo {author} {\bibfnamefont
  {W.~N.}\ \bibnamefont {Hardy}}, \bibinfo {author} {\bibfnamefont {D.~A.}\
  \bibnamefont {Bonn}}, \ and\ \bibinfo {author} {\bibfnamefont {M.-H.}\
  \bibnamefont {Julien}},\ }\href {https://doi.org/10.1038/ncomms3113}
  {\bibfield  {journal} {\bibinfo  {journal} {Nat. Comm.}\ }\textbf {\bibinfo
  {volume} {4}},\ \bibinfo {pages} {2113} (\bibinfo {year} {2013})}\BibitemShut
  {NoStop}%
\bibitem [{\citenamefont {Kim}\ \emph {et~al.}(2018)\citenamefont {Kim},
  \citenamefont {Souliou}, \citenamefont {Barber}, \citenamefont {Lefran{\c
  c}ois}, \citenamefont {Minola}, \citenamefont {Tortora}, \citenamefont
  {Heid}, \citenamefont {Nandi}, \citenamefont {Borzi}, \citenamefont
  {Garbarino}, \citenamefont {Bosak}, \citenamefont {Porras}, \citenamefont
  {Loew}, \citenamefont {König}, \citenamefont {Moll}, \citenamefont
  {Mackenzie}, \citenamefont {Keimer}, \citenamefont {Hicks},\ and\
  \citenamefont {Le~Tacon}}]{Kim2018}%
  \BibitemOpen
  \bibfield  {author} {\bibinfo {author} {\bibfnamefont {H.-H.}\ \bibnamefont
  {Kim}}, \bibinfo {author} {\bibfnamefont {S.~M.}\ \bibnamefont {Souliou}},
  \bibinfo {author} {\bibfnamefont {M.~E.}\ \bibnamefont {Barber}}, \bibinfo
  {author} {\bibfnamefont {E.}~\bibnamefont {Lefran{\c c}ois}}, \bibinfo
  {author} {\bibfnamefont {M.}~\bibnamefont {Minola}}, \bibinfo {author}
  {\bibfnamefont {M.}~\bibnamefont {Tortora}}, \bibinfo {author} {\bibfnamefont
  {R.}~\bibnamefont {Heid}}, \bibinfo {author} {\bibfnamefont {N.}~\bibnamefont
  {Nandi}}, \bibinfo {author} {\bibfnamefont {R.~A.}\ \bibnamefont {Borzi}},
  \bibinfo {author} {\bibfnamefont {G.}~\bibnamefont {Garbarino}}, \bibinfo
  {author} {\bibfnamefont {A.}~\bibnamefont {Bosak}}, \bibinfo {author}
  {\bibfnamefont {J.}~\bibnamefont {Porras}}, \bibinfo {author} {\bibfnamefont
  {T.}~\bibnamefont {Loew}}, \bibinfo {author} {\bibfnamefont {M.}~\bibnamefont
  {König}}, \bibinfo {author} {\bibfnamefont {P.~J.~W.}\ \bibnamefont {Moll}},
  \bibinfo {author} {\bibfnamefont {A.~P.}\ \bibnamefont {Mackenzie}}, \bibinfo
  {author} {\bibfnamefont {B.}~\bibnamefont {Keimer}}, \bibinfo {author}
  {\bibfnamefont {C.~W.}\ \bibnamefont {Hicks}}, \ and\ \bibinfo {author}
  {\bibfnamefont {M.}~\bibnamefont {Le~Tacon}},\ }\href
  {http://science.sciencemag.org/content/362/6418/1040.abstract} {\bibfield
  {journal} {\bibinfo  {journal} {Science}\ }\textbf {\bibinfo {volume}
  {362}},\ \bibinfo {pages} {1040} (\bibinfo {year} {2018})}\BibitemShut
  {NoStop}%
\bibitem [{\citenamefont {Doiron-Leyraud}\ \emph {et~al.}(2007)\citenamefont
  {Doiron-Leyraud}, \citenamefont {Proust}, \citenamefont {LeBoeuf},
  \citenamefont {Levallois}, \citenamefont {Bonnemaison}, \citenamefont
  {Liang}, \citenamefont {Bonn}, \citenamefont {Hardy},\ and\ \citenamefont
  {Taillefer}}]{Doiron-Leyraud2007}%
  \BibitemOpen
  \bibfield  {author} {\bibinfo {author} {\bibfnamefont {N.}~\bibnamefont
  {Doiron-Leyraud}}, \bibinfo {author} {\bibfnamefont {C.}~\bibnamefont
  {Proust}}, \bibinfo {author} {\bibfnamefont {D.}~\bibnamefont {LeBoeuf}},
  \bibinfo {author} {\bibfnamefont {J.}~\bibnamefont {Levallois}}, \bibinfo
  {author} {\bibfnamefont {J.-B.}\ \bibnamefont {Bonnemaison}}, \bibinfo
  {author} {\bibfnamefont {R.}~\bibnamefont {Liang}}, \bibinfo {author}
  {\bibfnamefont {D.~A.}\ \bibnamefont {Bonn}}, \bibinfo {author}
  {\bibfnamefont {W.~N.}\ \bibnamefont {Hardy}}, \ and\ \bibinfo {author}
  {\bibfnamefont {L.}~\bibnamefont {Taillefer}},\ }\href
  {http://dx.doi.org/10.1038/nature05872} {\bibfield  {journal} {\bibinfo
  {journal} {Nature}\ }\textbf {\bibinfo {volume} {447}},\ \bibinfo {pages}
  {565} (\bibinfo {year} {2007})}\BibitemShut {NoStop}%
\bibitem [{\citenamefont {Hussey}\ \emph {et~al.}(2003)\citenamefont {Hussey},
  \citenamefont {Abdel-Jawad}, \citenamefont {Carrington}, \citenamefont
  {Mackenzie},\ and\ \citenamefont {Balicas}}]{Hussey2003}%
  \BibitemOpen
  \bibfield  {author} {\bibinfo {author} {\bibfnamefont {N.~E.}\ \bibnamefont
  {Hussey}}, \bibinfo {author} {\bibfnamefont {M.}~\bibnamefont {Abdel-Jawad}},
  \bibinfo {author} {\bibfnamefont {A.}~\bibnamefont {Carrington}}, \bibinfo
  {author} {\bibfnamefont {A.~P.}\ \bibnamefont {Mackenzie}}, \ and\ \bibinfo
  {author} {\bibfnamefont {L.}~\bibnamefont {Balicas}},\ }\href
  {http://dx.doi.org/10.1038/nature01981} {\bibfield  {journal} {\bibinfo
  {journal} {Nature}\ }\textbf {\bibinfo {volume} {425}},\ \bibinfo {pages}
  {814} (\bibinfo {year} {2003})}\BibitemShut {NoStop}%
\bibitem [{\citenamefont {Vignolle}\ \emph {et~al.}(2008)\citenamefont
  {Vignolle}, \citenamefont {Carrington}, \citenamefont {Cooper}, \citenamefont
  {French}, \citenamefont {Mackenzie}, \citenamefont {Jaudet}, \citenamefont
  {Vignolles}, \citenamefont {Proust},\ and\ \citenamefont
  {Hussey}}]{Vignolle2008}%
  \BibitemOpen
  \bibfield  {author} {\bibinfo {author} {\bibfnamefont {B.}~\bibnamefont
  {Vignolle}}, \bibinfo {author} {\bibfnamefont {A.}~\bibnamefont
  {Carrington}}, \bibinfo {author} {\bibfnamefont {R.~A.}\ \bibnamefont
  {Cooper}}, \bibinfo {author} {\bibfnamefont {M.~M.~J.}\ \bibnamefont
  {French}}, \bibinfo {author} {\bibfnamefont {A.~P.}\ \bibnamefont
  {Mackenzie}}, \bibinfo {author} {\bibfnamefont {C.}~\bibnamefont {Jaudet}},
  \bibinfo {author} {\bibfnamefont {D.}~\bibnamefont {Vignolles}}, \bibinfo
  {author} {\bibfnamefont {C.}~\bibnamefont {Proust}}, \ and\ \bibinfo {author}
  {\bibfnamefont {N.~E.}\ \bibnamefont {Hussey}},\ }\href
  {http://dx.doi.org/10.1038/nature07323} {\bibfield  {journal} {\bibinfo
  {journal} {Nature}\ }\textbf {\bibinfo {volume} {455}},\ \bibinfo {pages}
  {952} (\bibinfo {year} {2008})}\BibitemShut {NoStop}%
\bibitem [{\citenamefont {Bari{\v s}i{\'c}}\ \emph {et~al.}(2013)\citenamefont
  {Bari{\v s}i{\'c}}, \citenamefont {Chan}, \citenamefont {Li}, \citenamefont
  {Yu}, \citenamefont {Zhao}, \citenamefont {Dressel}, \citenamefont
  {Smontara},\ and\ \citenamefont {Greven}}]{Barisic2013}%
  \BibitemOpen
  \bibfield  {author} {\bibinfo {author} {\bibfnamefont {N.}~\bibnamefont
  {Bari{\v s}i{\'c}}}, \bibinfo {author} {\bibfnamefont {M.~K.}\ \bibnamefont
  {Chan}}, \bibinfo {author} {\bibfnamefont {Y.}~\bibnamefont {Li}}, \bibinfo
  {author} {\bibfnamefont {G.}~\bibnamefont {Yu}}, \bibinfo {author}
  {\bibfnamefont {X.}~\bibnamefont {Zhao}}, \bibinfo {author} {\bibfnamefont
  {M.}~\bibnamefont {Dressel}}, \bibinfo {author} {\bibfnamefont
  {A.}~\bibnamefont {Smontara}}, \ and\ \bibinfo {author} {\bibfnamefont
  {M.}~\bibnamefont {Greven}},\ }\href {\doibase 10.1073/pnas.1301989110}
  {\bibfield  {journal} {\bibinfo  {journal} {Proc. Natl. Acad. Sci. U.S.A.}\
  }\textbf {\bibinfo {volume} {110}},\ \bibinfo {pages} {12235} (\bibinfo
  {year} {2013})}\BibitemShut {NoStop}%
\bibitem [{\citenamefont {Hossain}\ \emph {et~al.}(2008)\citenamefont
  {Hossain}, \citenamefont {Mottershead}, \citenamefont {Fournier},
  \citenamefont {Bostwick}, \citenamefont {McChesney}, \citenamefont
  {Rotenberg}, \citenamefont {Liang}, \citenamefont {Hardy}, \citenamefont
  {Sawatzky}, \citenamefont {Elfimov}, \citenamefont {Bonn},\ and\
  \citenamefont {Damascelli}}]{Hossain2008}%
  \BibitemOpen
  \bibfield  {author} {\bibinfo {author} {\bibfnamefont {M.~A.}\ \bibnamefont
  {Hossain}}, \bibinfo {author} {\bibfnamefont {J.~D.~F.}\ \bibnamefont
  {Mottershead}}, \bibinfo {author} {\bibfnamefont {D.}~\bibnamefont
  {Fournier}}, \bibinfo {author} {\bibfnamefont {A.}~\bibnamefont {Bostwick}},
  \bibinfo {author} {\bibfnamefont {J.~L.}\ \bibnamefont {McChesney}}, \bibinfo
  {author} {\bibfnamefont {E.}~\bibnamefont {Rotenberg}}, \bibinfo {author}
  {\bibfnamefont {R.}~\bibnamefont {Liang}}, \bibinfo {author} {\bibfnamefont
  {W.~N.}\ \bibnamefont {Hardy}}, \bibinfo {author} {\bibfnamefont {G.~A.}\
  \bibnamefont {Sawatzky}}, \bibinfo {author} {\bibfnamefont {I.~S.}\
  \bibnamefont {Elfimov}}, \bibinfo {author} {\bibfnamefont {D.~A.}\
  \bibnamefont {Bonn}}, \ and\ \bibinfo {author} {\bibfnamefont
  {A.}~\bibnamefont {Damascelli}},\ }\href {https://doi.org/10.1038/nphys998}
  {\bibfield  {journal} {\bibinfo  {journal} {Nat. Phys.}\ }\textbf {\bibinfo
  {volume} {4}},\ \bibinfo {pages} {527} (\bibinfo {year} {2008})}\BibitemShut
  {NoStop}%
\bibitem [{\citenamefont {Helm}\ \emph {et~al.}(2009)\citenamefont {Helm},
  \citenamefont {Kartsovnik}, \citenamefont {Bartkowiak}, \citenamefont
  {Bittner}, \citenamefont {Lambacher}, \citenamefont {Erb}, \citenamefont
  {Wosnitza},\ and\ \citenamefont {Gross}}]{Helm2009}%
  \BibitemOpen
  \bibfield  {author} {\bibinfo {author} {\bibfnamefont {T.}~\bibnamefont
  {Helm}}, \bibinfo {author} {\bibfnamefont {M.~V.}\ \bibnamefont
  {Kartsovnik}}, \bibinfo {author} {\bibfnamefont {M.}~\bibnamefont
  {Bartkowiak}}, \bibinfo {author} {\bibfnamefont {N.}~\bibnamefont {Bittner}},
  \bibinfo {author} {\bibfnamefont {M.}~\bibnamefont {Lambacher}}, \bibinfo
  {author} {\bibfnamefont {A.}~\bibnamefont {Erb}}, \bibinfo {author}
  {\bibfnamefont {J.}~\bibnamefont {Wosnitza}}, \ and\ \bibinfo {author}
  {\bibfnamefont {R.}~\bibnamefont {Gross}},\ }\href
  {https://link.aps.org/doi/10.1103/PhysRevLett.103.157002} {\bibfield
  {journal} {\bibinfo  {journal} {Phys. Rev. Lett.}\ }\textbf {\bibinfo
  {volume} {103}},\ \bibinfo {pages} {157002} (\bibinfo {year}
  {2009})}\BibitemShut {NoStop}%
\bibitem [{\citenamefont {Ramshaw}\ \emph {et~al.}(2015)\citenamefont
  {Ramshaw}, \citenamefont {Sebastian}, \citenamefont {McDonald}, \citenamefont
  {Day}, \citenamefont {Tan}, \citenamefont {Zhu}, \citenamefont {Betts},
  \citenamefont {Liang}, \citenamefont {Bonn}, \citenamefont {Hardy},\ and\
  \citenamefont {Harrison}}]{Ramshaw317}%
  \BibitemOpen
  \bibfield  {author} {\bibinfo {author} {\bibfnamefont {B.~J.}\ \bibnamefont
  {Ramshaw}}, \bibinfo {author} {\bibfnamefont {S.~E.}\ \bibnamefont
  {Sebastian}}, \bibinfo {author} {\bibfnamefont {R.~D.}\ \bibnamefont
  {McDonald}}, \bibinfo {author} {\bibfnamefont {J.}~\bibnamefont {Day}},
  \bibinfo {author} {\bibfnamefont {B.~S.}\ \bibnamefont {Tan}}, \bibinfo
  {author} {\bibfnamefont {Z.}~\bibnamefont {Zhu}}, \bibinfo {author}
  {\bibfnamefont {J.~B.}\ \bibnamefont {Betts}}, \bibinfo {author}
  {\bibfnamefont {R.}~\bibnamefont {Liang}}, \bibinfo {author} {\bibfnamefont
  {D.~A.}\ \bibnamefont {Bonn}}, \bibinfo {author} {\bibfnamefont {W.~N.}\
  \bibnamefont {Hardy}}, \ and\ \bibinfo {author} {\bibfnamefont
  {N.}~\bibnamefont {Harrison}},\ }\href {\doibase 10.1126/science.aaa4990}
  {\bibfield  {journal} {\bibinfo  {journal} {Science}\ }\textbf {\bibinfo
  {volume} {348}},\ \bibinfo {pages} {317} (\bibinfo {year}
  {2015})}\BibitemShut {NoStop}%
\bibitem [{\citenamefont {Grissonnanche}\ \emph {et~al.}(2014)\citenamefont
  {Grissonnanche}, \citenamefont {Cyr-Choini{\`e}re}, \citenamefont
  {Lalibert{\'e}}, \citenamefont {Ren{\'e}~de Cotret}, \citenamefont
  {Juneau-Fecteau}, \citenamefont {Dufour-Beaus{\'e}jour}, \citenamefont
  {Delage}, \citenamefont {LeBoeuf}, \citenamefont {Chang}, \citenamefont
  {Ramshaw}, \citenamefont {Bonn}, \citenamefont {Hardy}, \citenamefont
  {Liang}, \citenamefont {Adachi}, \citenamefont {Hussey}, \citenamefont
  {Vignolle}, \citenamefont {Proust}, \citenamefont {Sutherland}, \citenamefont
  {Kr{\"a}mer}, \citenamefont {Park}, \citenamefont {Graf}, \citenamefont
  {Doiron-Leyraud},\ and\ \citenamefont {Taillefer}}]{Grissonnanche2014}%
  \BibitemOpen
  \bibfield  {author} {\bibinfo {author} {\bibfnamefont {G.}~\bibnamefont
  {Grissonnanche}}, \bibinfo {author} {\bibfnamefont {O.}~\bibnamefont
  {Cyr-Choini{\`e}re}}, \bibinfo {author} {\bibfnamefont {F.}~\bibnamefont
  {Lalibert{\'e}}}, \bibinfo {author} {\bibfnamefont {S.}~\bibnamefont
  {Ren{\'e}~de Cotret}}, \bibinfo {author} {\bibfnamefont {A.}~\bibnamefont
  {Juneau-Fecteau}}, \bibinfo {author} {\bibfnamefont {S.}~\bibnamefont
  {Dufour-Beaus{\'e}jour}}, \bibinfo {author} {\bibfnamefont {M.~{\`E}.}\
  \bibnamefont {Delage}}, \bibinfo {author} {\bibfnamefont {D.}~\bibnamefont
  {LeBoeuf}}, \bibinfo {author} {\bibfnamefont {J.}~\bibnamefont {Chang}},
  \bibinfo {author} {\bibfnamefont {B.~J.}\ \bibnamefont {Ramshaw}}, \bibinfo
  {author} {\bibfnamefont {D.~A.}\ \bibnamefont {Bonn}}, \bibinfo {author}
  {\bibfnamefont {W.~N.}\ \bibnamefont {Hardy}}, \bibinfo {author}
  {\bibfnamefont {R.}~\bibnamefont {Liang}}, \bibinfo {author} {\bibfnamefont
  {S.}~\bibnamefont {Adachi}}, \bibinfo {author} {\bibfnamefont {N.~E.}\
  \bibnamefont {Hussey}}, \bibinfo {author} {\bibfnamefont {B.}~\bibnamefont
  {Vignolle}}, \bibinfo {author} {\bibfnamefont {C.}~\bibnamefont {Proust}},
  \bibinfo {author} {\bibfnamefont {M.}~\bibnamefont {Sutherland}}, \bibinfo
  {author} {\bibfnamefont {S.}~\bibnamefont {Kr{\"a}mer}}, \bibinfo {author}
  {\bibfnamefont {J.~H.}\ \bibnamefont {Park}}, \bibinfo {author}
  {\bibfnamefont {D.}~\bibnamefont {Graf}}, \bibinfo {author} {\bibfnamefont
  {N.}~\bibnamefont {Doiron-Leyraud}}, \ and\ \bibinfo {author} {\bibfnamefont
  {L.}~\bibnamefont {Taillefer}},\ }\href
  {http://dx.doi.org/10.1038/ncomms4280} {\bibfield  {journal} {\bibinfo
  {journal} {Nat. Comm.}\ }\textbf {\bibinfo {volume} {5}},\ \bibinfo {pages}
  {3280} (\bibinfo {year} {2014})}\BibitemShut {NoStop}%
\bibitem [{\citenamefont {Tallon}\ \emph {et~al.}(1997)\citenamefont {Tallon},
  \citenamefont {Bernhard}, \citenamefont {Williams},\ and\ \citenamefont
  {Loram}}]{Tallon1997}%
  \BibitemOpen
  \bibfield  {author} {\bibinfo {author} {\bibfnamefont {J.~L.}\ \bibnamefont
  {Tallon}}, \bibinfo {author} {\bibfnamefont {C.}~\bibnamefont {Bernhard}},
  \bibinfo {author} {\bibfnamefont {G.~V.~M.}\ \bibnamefont {Williams}}, \ and\
  \bibinfo {author} {\bibfnamefont {J.~W.}\ \bibnamefont {Loram}},\ }\href
  {https://link.aps.org/doi/10.1103/PhysRevLett.79.5294} {\bibfield  {journal}
  {\bibinfo  {journal} {Phys. Rev. Lett.}\ }\textbf {\bibinfo {volume} {79}},\
  \bibinfo {pages} {5294} (\bibinfo {year} {1997})}\BibitemShut {NoStop}%
\bibitem [{\citenamefont {Alireza}\ \emph {et~al.}(2017)\citenamefont
  {Alireza}, \citenamefont {Zhang}, \citenamefont {Guo}, \citenamefont
  {Porras}, \citenamefont {Loew}, \citenamefont {Hsu}, \citenamefont
  {Lonzarich}, \citenamefont {Le~Tacon}, \citenamefont {Keimer},\ and\
  \citenamefont {Sebastian}}]{Alireza2017}%
  \BibitemOpen
  \bibfield  {author} {\bibinfo {author} {\bibfnamefont {P.~L.}\ \bibnamefont
  {Alireza}}, \bibinfo {author} {\bibfnamefont {G.~H.}\ \bibnamefont {Zhang}},
  \bibinfo {author} {\bibfnamefont {W.}~\bibnamefont {Guo}}, \bibinfo {author}
  {\bibfnamefont {J.}~\bibnamefont {Porras}}, \bibinfo {author} {\bibfnamefont
  {T.}~\bibnamefont {Loew}}, \bibinfo {author} {\bibfnamefont {Y.-T.}\
  \bibnamefont {Hsu}}, \bibinfo {author} {\bibfnamefont {G.~G.}\ \bibnamefont
  {Lonzarich}}, \bibinfo {author} {\bibfnamefont {M.}~\bibnamefont {Le~Tacon}},
  \bibinfo {author} {\bibfnamefont {B.}~\bibnamefont {Keimer}}, \ and\ \bibinfo
  {author} {\bibfnamefont {S.~E.}\ \bibnamefont {Sebastian}},\ }\href {\doibase
  10.1103/PhysRevB.95.100505} {\bibfield  {journal} {\bibinfo  {journal} {Phys.
  Rev. B}\ }\textbf {\bibinfo {volume} {95}},\ \bibinfo {pages} {100505}
  (\bibinfo {year} {2017})}\BibitemShut {NoStop}%
\bibitem [{\citenamefont {Sadewasser}\ \emph {et~al.}(1997)\citenamefont
  {Sadewasser}, \citenamefont {Wang}, \citenamefont {Schilling}, \citenamefont
  {Zheng}, \citenamefont {Paulikas},\ and\ \citenamefont
  {Veal}}]{Sadewasser1997}%
  \BibitemOpen
  \bibfield  {author} {\bibinfo {author} {\bibfnamefont {S.}~\bibnamefont
  {Sadewasser}}, \bibinfo {author} {\bibfnamefont {Y.}~\bibnamefont {Wang}},
  \bibinfo {author} {\bibfnamefont {J.~S.}\ \bibnamefont {Schilling}}, \bibinfo
  {author} {\bibfnamefont {H.}~\bibnamefont {Zheng}}, \bibinfo {author}
  {\bibfnamefont {A.~P.}\ \bibnamefont {Paulikas}}, \ and\ \bibinfo {author}
  {\bibfnamefont {B.~W.}\ \bibnamefont {Veal}},\ }\href
  {https://link.aps.org/doi/10.1103/PhysRevB.56.14168} {\bibfield  {journal}
  {\bibinfo  {journal} {Phys. Rev. B}\ }\textbf {\bibinfo {volume} {56}},\
  \bibinfo {pages} {14168} (\bibinfo {year} {1997})}\BibitemShut {NoStop}%
\bibitem [{\citenamefont {Tozer}\ \emph {et~al.}(1987)\citenamefont {Tozer},
  \citenamefont {Kleinsasser}, \citenamefont {Penney}, \citenamefont {Kaiser},\
  and\ \citenamefont {Holtzberg}}]{Tozer1987}%
  \BibitemOpen
  \bibfield  {author} {\bibinfo {author} {\bibfnamefont {S.~W.}\ \bibnamefont
  {Tozer}}, \bibinfo {author} {\bibfnamefont {A.~W.}\ \bibnamefont
  {Kleinsasser}}, \bibinfo {author} {\bibfnamefont {T.}~\bibnamefont {Penney}},
  \bibinfo {author} {\bibfnamefont {D.}~\bibnamefont {Kaiser}}, \ and\ \bibinfo
  {author} {\bibfnamefont {F.}~\bibnamefont {Holtzberg}},\ }\href
  {https://link.aps.org/doi/10.1103/PhysRevLett.59.1768} {\bibfield  {journal}
  {\bibinfo  {journal} {Phys. Rev. Lett.}\ }\textbf {\bibinfo {volume} {59}},\
  \bibinfo {pages} {1768} (\bibinfo {year} {1987})}\BibitemShut {NoStop}%
\bibitem [{\citenamefont {Tozer}\ \emph {et~al.}(1993)\citenamefont {Tozer},
  \citenamefont {Koston},\ and\ \citenamefont {McCarron~III}}]{Tozer1993}%
  \BibitemOpen
  \bibfield  {author} {\bibinfo {author} {\bibfnamefont {S.~W.}\ \bibnamefont
  {Tozer}}, \bibinfo {author} {\bibfnamefont {J.~L.}\ \bibnamefont {Koston}}, \
  and\ \bibinfo {author} {\bibfnamefont {E.~M.}\ \bibnamefont {McCarron~III}},\
  }\href {https://link.aps.org/doi/10.1103/PhysRevB.47.8089} {\bibfield
  {journal} {\bibinfo  {journal} {Phys. Rev. B}\ }\textbf {\bibinfo {volume}
  {47}},\ \bibinfo {pages} {8089} (\bibinfo {year} {1993})}\BibitemShut
  {NoStop}%
\bibitem [{\citenamefont {Graf}\ \emph {et~al.}(2011)\citenamefont {Graf},
  \citenamefont {Stillwell}, \citenamefont {Purcell},\ and\ \citenamefont
  {Tozer}}]{Graf2011}%
  \BibitemOpen
  \bibfield  {author} {\bibinfo {author} {\bibfnamefont {D.~E.}\ \bibnamefont
  {Graf}}, \bibinfo {author} {\bibfnamefont {R.~L.}\ \bibnamefont {Stillwell}},
  \bibinfo {author} {\bibfnamefont {K.~M.}\ \bibnamefont {Purcell}}, \ and\
  \bibinfo {author} {\bibfnamefont {S.~W.}\ \bibnamefont {Tozer}},\ }\href
  {\doibase 10.1080/08957959.2011.633909} {\bibfield  {journal} {\bibinfo
  {journal} {High Pressure Research}\ }\textbf {\bibinfo {volume} {31}},\
  \bibinfo {pages} {533} (\bibinfo {year} {2011})}\BibitemShut {NoStop}%
\bibitem [{\citenamefont {Savitzky}\ and\ \citenamefont
  {Golay}(1964)}]{Savitzky1964}%
  \BibitemOpen
  \bibfield  {author} {\bibinfo {author} {\bibfnamefont {A.}~\bibnamefont
  {Savitzky}}\ and\ \bibinfo {author} {\bibfnamefont {M.~J.~E.}\ \bibnamefont
  {Golay}},\ }\href {\doibase 10.1021/ac60214a047} {\bibfield  {journal}
  {\bibinfo  {journal} {Analytical Chemistry}\ }\textbf {\bibinfo {volume}
  {36}},\ \bibinfo {pages} {1627} (\bibinfo {year} {1964})}\BibitemShut
  {NoStop}%
\bibitem [{\citenamefont {{Schoenberg}}(2009)}]{Schoenberg2009}%
  \BibitemOpen
  \bibfield  {author} {\bibinfo {author} {\bibfnamefont {D.}~\bibnamefont
  {{Schoenberg}}},\ }\href@noop {} {\emph {\bibinfo {title} {Magnetic
  Oscillations in Metals}}}\ (\bibinfo  {publisher} {Cambridge University
  Press},\ \bibinfo {year} {2009})\BibitemShut {NoStop}%
\bibitem [{\citenamefont {Tateiwa}\ and\ \citenamefont
  {Haga}(2009)}]{Tateiwa2009}%
  \BibitemOpen
  \bibfield  {author} {\bibinfo {author} {\bibfnamefont {N.}~\bibnamefont
  {Tateiwa}}\ and\ \bibinfo {author} {\bibfnamefont {Y.}~\bibnamefont {Haga}},\
  }\href {\doibase 10.1063/1.3265992} {\bibfield  {journal} {\bibinfo
  {journal} {Rev. Sci. Instr.}\ }\textbf {\bibinfo {volume} {80}},\ \bibinfo
  {pages} {123901} (\bibinfo {year} {2009})}\BibitemShut {NoStop}%
\bibitem [{\citenamefont {Sadewasser}\ \emph {et~al.}(2000)\citenamefont
  {Sadewasser}, \citenamefont {Schilling}, \citenamefont {Paulikas},\ and\
  \citenamefont {Veal}}]{Sadewasser2000}%
  \BibitemOpen
  \bibfield  {author} {\bibinfo {author} {\bibfnamefont {S.}~\bibnamefont
  {Sadewasser}}, \bibinfo {author} {\bibfnamefont {J.~S.}\ \bibnamefont
  {Schilling}}, \bibinfo {author} {\bibfnamefont {A.~P.}\ \bibnamefont
  {Paulikas}}, \ and\ \bibinfo {author} {\bibfnamefont {B.~W.}\ \bibnamefont
  {Veal}},\ }\href {\doibase 10.1103/PhysRevB.61.741} {\bibfield  {journal}
  {\bibinfo  {journal} {Phys. Rev. B}\ }\textbf {\bibinfo {volume} {61}},\
  \bibinfo {pages} {741} (\bibinfo {year} {2000})}\BibitemShut {NoStop}%
\bibitem [{\citenamefont {Cyr-Choini{\`e}re}\ \emph {et~al.}(2018)\citenamefont
  {Cyr-Choini{\`e}re}, \citenamefont {LeBoeuf}, \citenamefont {Badoux},
  \citenamefont {Dufour-Beaus{\'e}jour}, \citenamefont {Bonn}, \citenamefont
  {Hardy}, \citenamefont {Liang}, \citenamefont {Graf}, \citenamefont
  {Doiron-Leyraud},\ and\ \citenamefont {Taillefer}}]{Cyr-Choiniere2018}%
  \BibitemOpen
  \bibfield  {author} {\bibinfo {author} {\bibfnamefont {O.}~\bibnamefont
  {Cyr-Choini{\`e}re}}, \bibinfo {author} {\bibfnamefont {D.}~\bibnamefont
  {LeBoeuf}}, \bibinfo {author} {\bibfnamefont {S.}~\bibnamefont {Badoux}},
  \bibinfo {author} {\bibfnamefont {S.}~\bibnamefont {Dufour-Beaus{\'e}jour}},
  \bibinfo {author} {\bibfnamefont {D.~A.}\ \bibnamefont {Bonn}}, \bibinfo
  {author} {\bibfnamefont {W.~N.}\ \bibnamefont {Hardy}}, \bibinfo {author}
  {\bibfnamefont {R.}~\bibnamefont {Liang}}, \bibinfo {author} {\bibfnamefont
  {D.}~\bibnamefont {Graf}}, \bibinfo {author} {\bibfnamefont {N.}~\bibnamefont
  {Doiron-Leyraud}}, \ and\ \bibinfo {author} {\bibfnamefont {L.}~\bibnamefont
  {Taillefer}},\ }\href {https://link.aps.org/doi/10.1103/PhysRevB.98.064513}
  {\bibfield  {journal} {\bibinfo  {journal} {Phys. Rev. B}\ }\textbf {\bibinfo
  {volume} {98}},\ \bibinfo {pages} {064513} (\bibinfo {year}
  {2018})}\BibitemShut {NoStop}%
\bibitem [{\citenamefont {Jurkutat}\ \emph {et~al.}(2023)\citenamefont
  {Jurkutat}, \citenamefont {Kattinger}, \citenamefont {Tsankov}, \citenamefont
  {Reznicek}, \citenamefont {Erb},\ and\ \citenamefont {Haase}}]{Jurkutat2023}%
  \BibitemOpen
  \bibfield  {author} {\bibinfo {author} {\bibfnamefont {M.}~\bibnamefont
  {Jurkutat}}, \bibinfo {author} {\bibfnamefont {C.}~\bibnamefont {Kattinger}},
  \bibinfo {author} {\bibfnamefont {S.}~\bibnamefont {Tsankov}}, \bibinfo
  {author} {\bibfnamefont {R.}~\bibnamefont {Reznicek}}, \bibinfo {author}
  {\bibfnamefont {A.}~\bibnamefont {Erb}}, \ and\ \bibinfo {author}
  {\bibfnamefont {J.}~\bibnamefont {Haase}},\ }\href {\doibase
  10.1073/pnas.2215458120} {\bibfield  {journal} {\bibinfo  {journal} {Proc.
  Natl. Acad. Sci. U.S.A.}\ }\textbf {\bibinfo {volume} {120}},\ \bibinfo
  {pages} {e2215458120} (\bibinfo {year} {2023})}\BibitemShut {NoStop}%
\bibitem [{\citenamefont {Kawasaki}\ \emph {et~al.}(2017)\citenamefont
  {Kawasaki}, \citenamefont {Li}, \citenamefont {Kitahashi}, \citenamefont
  {Lin}, \citenamefont {Kuhns}, \citenamefont {Reyes},\ and\ \citenamefont
  {Zheng}}]{Kawasaki2017}%
  \BibitemOpen
  \bibfield  {author} {\bibinfo {author} {\bibfnamefont {S.}~\bibnamefont
  {Kawasaki}}, \bibinfo {author} {\bibfnamefont {Z.}~\bibnamefont {Li}},
  \bibinfo {author} {\bibfnamefont {M.}~\bibnamefont {Kitahashi}}, \bibinfo
  {author} {\bibfnamefont {C.~T.}\ \bibnamefont {Lin}}, \bibinfo {author}
  {\bibfnamefont {P.~L.}\ \bibnamefont {Kuhns}}, \bibinfo {author}
  {\bibfnamefont {A.~P.}\ \bibnamefont {Reyes}}, \ and\ \bibinfo {author}
  {\bibfnamefont {G.-q.}\ \bibnamefont {Zheng}},\ }\href
  {https://doi.org/10.1038/s41467-017-01465-9} {\bibfield  {journal} {\bibinfo
  {journal} {Nat. Comm.}\ }\textbf {\bibinfo {volume} {8}},\ \bibinfo {pages}
  {1267} (\bibinfo {year} {2017})}\BibitemShut {NoStop}%
\bibitem [{\citenamefont {Olsen}\ \emph {et~al.}(1991)\citenamefont {Olsen},
  \citenamefont {Steenstrup}, \citenamefont {Gerward},\ and\ \citenamefont
  {Sundqvist}}]{Olsen1991}%
  \BibitemOpen
  \bibfield  {author} {\bibinfo {author} {\bibfnamefont {J.~S.}\ \bibnamefont
  {Olsen}}, \bibinfo {author} {\bibfnamefont {S.}~\bibnamefont {Steenstrup}},
  \bibinfo {author} {\bibfnamefont {L.}~\bibnamefont {Gerward}}, \ and\
  \bibinfo {author} {\bibfnamefont {B.}~\bibnamefont {Sundqvist}},\ }\href
  {\doibase 10.1088/0031-8949/44/2/017} {\bibfield  {journal} {\bibinfo
  {journal} {Phys. Scr.}\ }\textbf {\bibinfo {volume} {44}},\ \bibinfo {pages}
  {211} (\bibinfo {year} {1991})}\BibitemShut {NoStop}%
\bibitem [{\citenamefont {Ludwig}\ \emph {et~al.}(1992)\citenamefont {Ludwig},
  \citenamefont {Fietz},\ and\ \citenamefont {Wühl}}]{Ludwig1992}%
  \BibitemOpen
  \bibfield  {author} {\bibinfo {author} {\bibfnamefont {H.~A.}\ \bibnamefont
  {Ludwig}}, \bibinfo {author} {\bibfnamefont {W.~H.}\ \bibnamefont {Fietz}}, \
  and\ \bibinfo {author} {\bibfnamefont {H.}~\bibnamefont {Wühl}},\ }\href
  {\doibase 10.1016/0921-4534(92)90244-7} {\bibfield  {journal} {\bibinfo
  {journal} {Physica C Supercond}\ }\textbf {\bibinfo {volume} {197}},\
  \bibinfo {pages} {113} (\bibinfo {year} {1992})}\BibitemShut {NoStop}%
\bibitem [{\citenamefont {Abdel-Jawad}\ \emph {et~al.}(2006)\citenamefont
  {Abdel-Jawad}, \citenamefont {Kennett}, \citenamefont {Balicas},
  \citenamefont {Carrington}, \citenamefont {Mackenzie}, \citenamefont
  {McKenzie},\ and\ \citenamefont {Hussey}}]{Abdel-Jawad2006}%
  \BibitemOpen
  \bibfield  {author} {\bibinfo {author} {\bibfnamefont {M.}~\bibnamefont
  {Abdel-Jawad}}, \bibinfo {author} {\bibfnamefont {M.~P.}\ \bibnamefont
  {Kennett}}, \bibinfo {author} {\bibfnamefont {L.}~\bibnamefont {Balicas}},
  \bibinfo {author} {\bibfnamefont {A.}~\bibnamefont {Carrington}}, \bibinfo
  {author} {\bibfnamefont {A.~P.}\ \bibnamefont {Mackenzie}}, \bibinfo {author}
  {\bibfnamefont {R.~H.}\ \bibnamefont {McKenzie}}, \ and\ \bibinfo {author}
  {\bibfnamefont {N.~E.}\ \bibnamefont {Hussey}},\ }\href
  {http://dx.doi.org/10.1038/nphys449} {\bibfield  {journal} {\bibinfo
  {journal} {Nat. Phys.}\ }\textbf {\bibinfo {volume} {2}},\ \bibinfo {pages}
  {821} (\bibinfo {year} {2006})}\BibitemShut {NoStop}%
\bibitem [{\citenamefont {Damascelli}\ \emph {et~al.}(2001)\citenamefont
  {Damascelli}, \citenamefont {Lu},\ and\ \citenamefont
  {Shen}}]{Damascelli2001}%
  \BibitemOpen
  \bibfield  {author} {\bibinfo {author} {\bibfnamefont {A.}~\bibnamefont
  {Damascelli}}, \bibinfo {author} {\bibfnamefont {D.~H.}\ \bibnamefont {Lu}},
  \ and\ \bibinfo {author} {\bibfnamefont {Z.-X.}\ \bibnamefont {Shen}},\
  }\href {\doibase 10.1016/s0368-2048(01)00264-x} {\bibfield  {journal}
  {\bibinfo  {journal} {Journal of Electron Spectroscopy and Related
  Phenomena}\ }\textbf {\bibinfo {volume} {117}},\ \bibinfo {pages} {165}
  (\bibinfo {year} {2001})}\BibitemShut {NoStop}%
\bibitem [{\citenamefont {Plat{\'e}}\ \emph {et~al.}(2005)\citenamefont
  {Plat{\'e}}, \citenamefont {Mottershead}, \citenamefont {Elfimov},
  \citenamefont {Peets}, \citenamefont {Liang}, \citenamefont {Bonn},
  \citenamefont {Hardy}, \citenamefont {Chiuzbaian}, \citenamefont {Falub},
  \citenamefont {Shi}, \citenamefont {Patthey},\ and\ \citenamefont
  {Damascelli}}]{Plate2005}%
  \BibitemOpen
  \bibfield  {author} {\bibinfo {author} {\bibfnamefont {M.}~\bibnamefont
  {Plat{\'e}}}, \bibinfo {author} {\bibfnamefont {J.~D.~F.}\ \bibnamefont
  {Mottershead}}, \bibinfo {author} {\bibfnamefont {I.~S.}\ \bibnamefont
  {Elfimov}}, \bibinfo {author} {\bibfnamefont {D.~C.}\ \bibnamefont {Peets}},
  \bibinfo {author} {\bibfnamefont {R.}~\bibnamefont {Liang}}, \bibinfo
  {author} {\bibfnamefont {D.~A.}\ \bibnamefont {Bonn}}, \bibinfo {author}
  {\bibfnamefont {W.~N.}\ \bibnamefont {Hardy}}, \bibinfo {author}
  {\bibfnamefont {S.}~\bibnamefont {Chiuzbaian}}, \bibinfo {author}
  {\bibfnamefont {M.}~\bibnamefont {Falub}}, \bibinfo {author} {\bibfnamefont
  {M.}~\bibnamefont {Shi}}, \bibinfo {author} {\bibfnamefont {L.}~\bibnamefont
  {Patthey}}, \ and\ \bibinfo {author} {\bibfnamefont {A.}~\bibnamefont
  {Damascelli}},\ }\href
  {https://link.aps.org/doi/10.1103/PhysRevLett.95.077001} {\bibfield
  {journal} {\bibinfo  {journal} {Phys. Rev. Lett.}\ }\textbf {\bibinfo
  {volume} {95}},\ \bibinfo {pages} {077001} (\bibinfo {year}
  {2005})}\BibitemShut {NoStop}%
\bibitem [{\citenamefont {Vignolle}\ \emph {et~al.}(2011)\citenamefont
  {Vignolle}, \citenamefont {Vignolles}, \citenamefont {LeBoeuf}, \citenamefont
  {Lepault}, \citenamefont {Ramshaw}, \citenamefont {Liang}, \citenamefont
  {Bonn}, \citenamefont {Hardy}, \citenamefont {Doiron-Leyraud}, \citenamefont
  {Carrington}, \citenamefont {Hussey}, \citenamefont {Taillefer},\ and\
  \citenamefont {Proust}}]{Vignolle2011}%
  \BibitemOpen
  \bibfield  {author} {\bibinfo {author} {\bibfnamefont {B.}~\bibnamefont
  {Vignolle}}, \bibinfo {author} {\bibfnamefont {D.}~\bibnamefont {Vignolles}},
  \bibinfo {author} {\bibfnamefont {D.}~\bibnamefont {LeBoeuf}}, \bibinfo
  {author} {\bibfnamefont {S.}~\bibnamefont {Lepault}}, \bibinfo {author}
  {\bibfnamefont {B.}~\bibnamefont {Ramshaw}}, \bibinfo {author} {\bibfnamefont
  {R.}~\bibnamefont {Liang}}, \bibinfo {author} {\bibfnamefont {D.~A.}\
  \bibnamefont {Bonn}}, \bibinfo {author} {\bibfnamefont {W.~N.}\ \bibnamefont
  {Hardy}}, \bibinfo {author} {\bibfnamefont {N.}~\bibnamefont
  {Doiron-Leyraud}}, \bibinfo {author} {\bibfnamefont {A.}~\bibnamefont
  {Carrington}}, \bibinfo {author} {\bibfnamefont {N.~E.}\ \bibnamefont
  {Hussey}}, \bibinfo {author} {\bibfnamefont {L.}~\bibnamefont {Taillefer}}, \
  and\ \bibinfo {author} {\bibfnamefont {C.}~\bibnamefont {Proust}},\ }\href
  {http://www.sciencedirect.com/science/article/pii/S1631070511000946}
  {\bibfield  {journal} {\bibinfo  {journal} {Comptes Rendus Physique}\
  }\textbf {\bibinfo {volume} {12}},\ \bibinfo {pages} {446} (\bibinfo {year}
  {2011})}\BibitemShut {NoStop}%
\bibitem [{\citenamefont {Van~Degrift}(2019)}]{VanDegrift2019}%
  \BibitemOpen
  \bibfield  {author} {\bibinfo {author} {\bibfnamefont {C.~T.}\ \bibnamefont
  {Van~Degrift}},\ }\href {\doibase 10.1063/1.1134272} {\bibfield  {journal}
  {\bibinfo  {journal} {Rev. Sci. Instr.}\ }\textbf {\bibinfo {volume} {46}},\
  \bibinfo {pages} {599} (\bibinfo {year} {2019})}\BibitemShut {NoStop}%
\bibitem [{\citenamefont {Coniglio}\ \emph {et~al.}(2010)\citenamefont
  {Coniglio}, \citenamefont {Winter}, \citenamefont {Rea}, \citenamefont
  {Cho},\ and\ \citenamefont {Agosta}}]{Coniglio2010}%
  \BibitemOpen
  \bibfield  {author} {\bibinfo {author} {\bibfnamefont {W.~A.}\ \bibnamefont
  {Coniglio}}, \bibinfo {author} {\bibfnamefont {L.~E.}\ \bibnamefont
  {Winter}}, \bibinfo {author} {\bibfnamefont {C.}~\bibnamefont {Rea}},
  \bibinfo {author} {\bibfnamefont {K.}~\bibnamefont {Cho}}, \ and\ \bibinfo
  {author} {\bibfnamefont {C.}~\bibnamefont {Agosta}},\ }\href {\doibase
  1003.5233v1} {\bibfield  {journal} {\bibinfo  {journal} {arXiv}\ } (\bibinfo
  {year} {2010}),\ 1003.5233v1}\BibitemShut {NoStop}%
\bibitem [{\citenamefont {Zherlitsyn}\ \emph {et~al.}(2012)\citenamefont
  {Zherlitsyn}, \citenamefont {Wustmann}, \citenamefont {Herrmannsdorfer},\
  and\ \citenamefont {Wosnitza}}]{Zherlitsyn2012}%
  \BibitemOpen
  \bibfield  {author} {\bibinfo {author} {\bibfnamefont {S.}~\bibnamefont
  {Zherlitsyn}}, \bibinfo {author} {\bibfnamefont {B.}~\bibnamefont
  {Wustmann}}, \bibinfo {author} {\bibfnamefont {T.}~\bibnamefont
  {Herrmannsdorfer}}, \ and\ \bibinfo {author} {\bibfnamefont {J.}~\bibnamefont
  {Wosnitza}},\ }\href {\doibase 10.1109/TASC.2012.2182975} {\bibfield
  {journal} {\bibinfo  {journal} {{IEEE} Transactions on Applied
  Superconductivity}\ }\textbf {\bibinfo {volume} {22}},\ \bibinfo {pages}
  {4300603} (\bibinfo {year} {2012})}\BibitemShut {NoStop}%
\bibitem [{\citenamefont {Forman}\ \emph {et~al.}(1972)\citenamefont {Forman},
  \citenamefont {Piermarini}, \citenamefont {Barnett},\ and\ \citenamefont
  {Block}}]{Forman1972}%
  \BibitemOpen
  \bibfield  {author} {\bibinfo {author} {\bibfnamefont {R.~A.}\ \bibnamefont
  {Forman}}, \bibinfo {author} {\bibfnamefont {G.~J.}\ \bibnamefont
  {Piermarini}}, \bibinfo {author} {\bibfnamefont {J.~D.}\ \bibnamefont
  {Barnett}}, \ and\ \bibinfo {author} {\bibfnamefont {S.}~\bibnamefont
  {Block}},\ }\href
  {http://science.sciencemag.org/content/176/4032/284.abstract} {\bibfield
  {journal} {\bibinfo  {journal} {Science}\ }\textbf {\bibinfo {volume}
  {176}},\ \bibinfo {pages} {284} (\bibinfo {year} {1972})}\BibitemShut
  {NoStop}%
\bibitem [{\citenamefont {Klotz}\ \emph {et~al.}(2009)\citenamefont {Klotz},
  \citenamefont {Chervin}, \citenamefont {Munsch},\ and\ \citenamefont
  {Marchand}}]{Klotz2009}%
  \BibitemOpen
  \bibfield  {author} {\bibinfo {author} {\bibfnamefont {S.}~\bibnamefont
  {Klotz}}, \bibinfo {author} {\bibfnamefont {J.-C.}\ \bibnamefont {Chervin}},
  \bibinfo {author} {\bibfnamefont {P.}~\bibnamefont {Munsch}}, \ and\ \bibinfo
  {author} {\bibfnamefont {G.~L.}\ \bibnamefont {Marchand}},\ }\href
  {http://stacks.iop.org/0022-3727/42/i=7/a=075413} {\bibfield  {journal}
  {\bibinfo  {journal} {Journal of Physics D: Applied Physics}\ }\textbf
  {\bibinfo {volume} {42}},\ \bibinfo {pages} {075413} (\bibinfo {year}
  {2009})}\BibitemShut {NoStop}%
\end{thebibliography}%

\renewcommand{\figurename}{Supplementary Figure}
\setcounter{figure}{0}
\section{Supplementary Information}
\subparagraph{Tuned LC tank circuit}
The tunnel-diode-oscillator (TDO) technique is a RLC tank circuit driven by a tunnel diode oscillator as described in Ref. \cite{VanDegrift2019} and  \cite{Coniglio2010}. The inductor coil inside the sample space of the pressure cell is shown in Sup.  Fig.  \ref{SI3_photo2_setup.fig}. The diode card,  consisting of the tunnel diode, capacitors and resistor,  is located at cryogenic temperatures 3 cm above the sample. This reduces the capacitance of the circuit between the diode and the sample,  ensuring a larger signal,  and allows us to drive the circuit at a higher frequency so that the change in frequency is larger.  The resonant frequency measures the skin depth for a metal and the penetration depth of a superconductor. The portion of the sample actually measured is proportional to the inverse resonant frequency and the resistivity of the sample,  according to:  
\begin{align}
\delta &= \sqrt{\dfrac{\rho}{\pi f\mu}}, \label{skindepth} 
\end{align}
where $\delta$ is the RF skin depth probed, $\rho$ the resistivity,  $f$ the frequency,  and $\mu$ the permeability. Taking $\mu = \mu_{0}$ for a conductor,  $f$ of the order of 500 MHz and $\rho$ on the order of 0.1 m$\Omega$.cm above $T_{c}$ for a hole concentration p of 0.09 \cite{Barisic2013}, we obtain a skin depth on the order of 20 $\mu$m. The typical sample dimensions are 100x100x20 $\mu$m$^{3}$. Therefore, we are able to probe the entire thickness of the YBCO sample, while maintaining efficient coupling between the skin depth and inductor.  It must be noted that the power dissipated by the TDO circuit at the sample at base $^{3}$He temperatures,  combined with the low thermal conductivity in a pressure cell due to a glassy pressure medium,  led to discrepancies between the temperature measured and the actual sample temperature.  The power dissipated by the TDO card has been estimated to be less than 10 $\mu$ W which would not affect the LK fits,  that rely more heavily on the higher temperature points. \\

\begin{figure}[t]
\begin{center}	
	\includegraphics[scale=0.8]{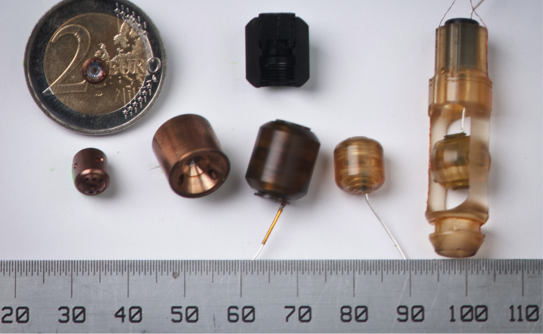}
		\caption{Various turnbuckle DACs, from left to right: 6 mm diameter DAC for MPMS work,  large-angle DAC for optical and x-ray studies, EC17 model made of black Parmax (the metallic version of which was used for our DC field studies), EC15b model made of Parmax, and the EC15b cell in a plastic pulsed-field rotator.}
		\label{SI1_photo1.fig}	
\end{center}		
\end{figure}

\subparagraph{High pressure}
This high pressure study of YBCO6.5 was performed using custom-made diamond anvil cells (DACs) based on the turnbuckle design as described in Ref. \cite{Graf2011}. Various cells that make use of that concept are shown in Sup.  Fig.  \ref{SI1_photo1.fig}.  Non-magnetic metallic DACs (BeCu C17200) were used for the static DC magnetic field measurements.  Measurements in pulsed magnetic fields at $^{3}$He temperatures require additional precautions to limit the vibrations and heating generated by the high $dB/dt$. The rapidly increasing field on the upsweep of a 70 T coil is of the order of 2400 T/s, and, respectively, 770 and 19000 T/s for outsert and insert of the 85 T duplex coil at HLD-EMFL \cite{Zherlitsyn2012}. To limit these deleterious effects, our DACs and rotators for pulsed-field studies are made of the plastic Parmax (SRP120) \cite{Graf2011}, a high strength self-reinforced polymer; the $^{3}$He and $^{4}$He cryostat tails are made from spiral-wound 75 $\mu$m thick G10 sheets.  Any heating would result in an offset between the up- and downsweep traces and a diminished amplitude of the QO.  As shown in Fig.  \ref{pulsed.fig} by the near perfect overlap of the upsweep and downsweep traces, negligible heating of the sample occurs during a pulse at $^{3}$He base temperature. \\
As shown in Sup.  Fig.  \ref{SI1_photo1.fig},  our DACs have been designed to be rotated in the constrained space of magnet bores at cryogenic temperatures (32 mm for DC field; 15.5 mm for pulsed magnets), which enables us to carefully align our samples in field and carry out angular dependence studies. \\
Supplementary Figure \ref{SI3_photo2_setup.fig} shows a typical TDO DAC set up. The gaskets used are composed of two parts: an insulating blend of diamond powder and epoxy compressed between the anvils, and a ring epoxied around the culet to hold that mix. That ring is made of either radially-wound Zylon fiber for the plastic cells or  a stainless-steel 304 gasket for the metallic cells.  The stainless-steel gaskets were pre-indented, and the indented region drilled out and filled with the diamond powder and epoxy mix.  Various culet sizes were used, from 450 to 800 $\mu$m, depending on the desired maximum pressure.  For pressures below 10 GPa, the typical geometry of our hand-wound TDO coils is a three-turn coil,  with an ID of 100-150 $\mu$m made using AWG56 (12 $\mu$m) or AWG59 (9 $\mu$m) copper wire with polynylon insulation.  For pressures greater than 10 GPa,  2-turn coils were used to have more height available in the sample space to compress the gasket.  Reducing the number of turns in the coil reduces the amplitude of the TDO signal (see results in Fig. \ref{fig7.fig}).  \\
Ruby spheres (RSA Le Rubis) are placed in the sample space of the DAC.  A 532 nm laser was used to excite the R1 and R2 lines of the ruby.  This enabled systematic measurements of the pressure at low temperature with an optical fiber placed against the upper diamond anvil.  For $T_{c}$ measurements, the pressure is also measured at the onset temperature of superconductivity since the pressure changes with temperature.  An additional optical fiber with ambient ruby is located in close proximity to the DAC to account for the temperature dependence of the ruby fluorescence.  Laser power is minimized to avoid any heating of the ruby during the pressure measurement. The value of the pressure is given by $p$ = (cell R1 - ambient R1) (nm) / 0.365 (nm/GPa) \cite{Forman1972}.  Several pressure media were utilized; glycerin,  Daphne 7474, and a mixture of 4:1::methanol:ethanol for pressures greater than 5 GPa.  Different pressure media have different hydrostatic limits  (see \cite{Tateiwa2009},  \cite{Klotz2009}), and these decrease with decreasing temperatures.  The hydrostatic limit for 4:1::methanol:ethanol is 10 GPa at room temperature,  which means that our pressures greater than 18 GPa are quasi-hydrostatic. The hydrostaticity level  within a cell under pressure is monitored by  observing the FWHM of the ruby fluorescence lines.  Non-hydrostatic conditions strain the sample and create a condition similar to scattering which can suppress QOs. \\

\begin{figure}[t]
\begin{center}	
	\includegraphics[scale=0.3]{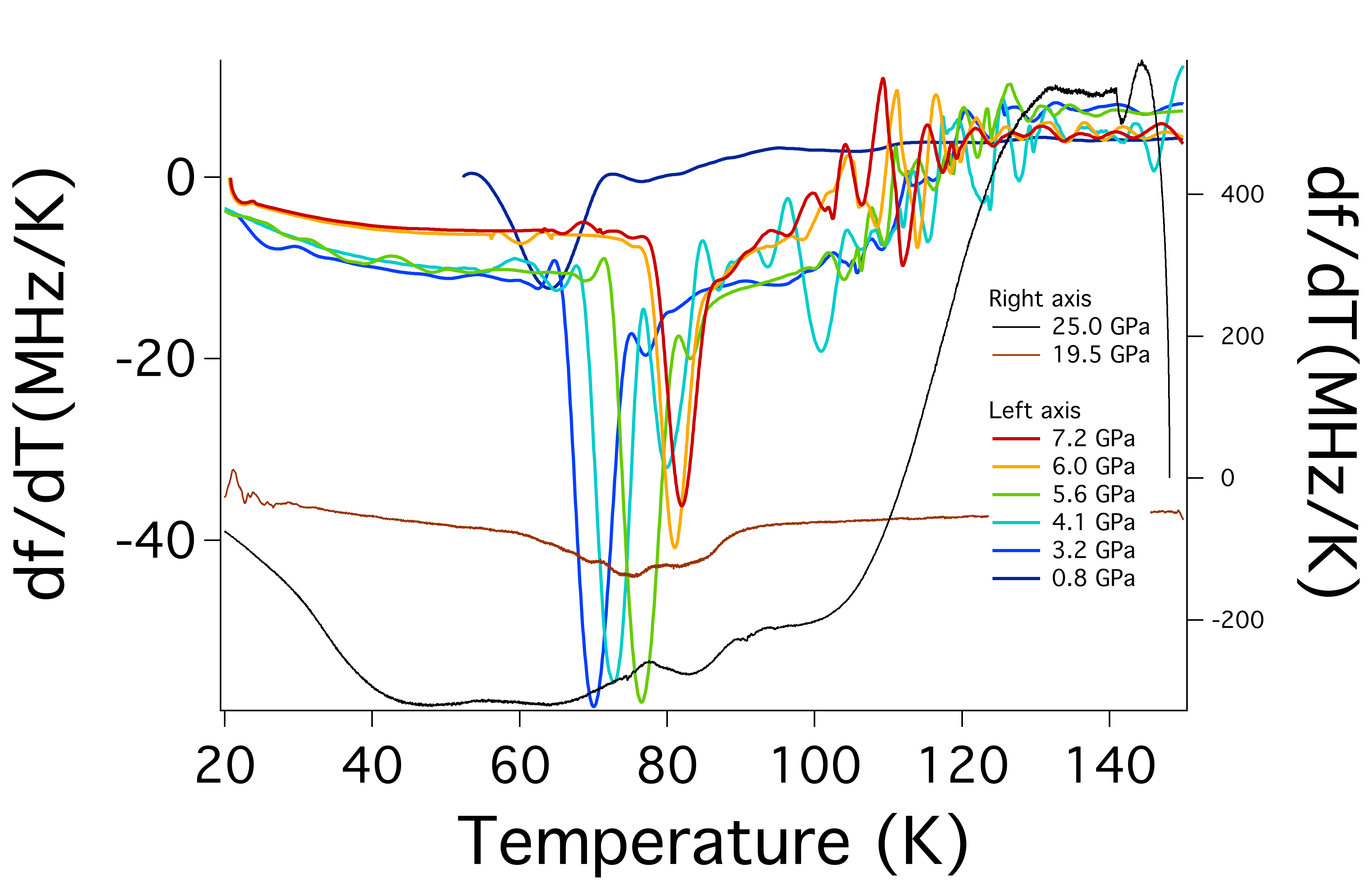}
		\caption{Derivative of the TDO frequency of the data in Fig.\ref{Tcvsp.fig}. The minima around 80 K are the $T_{c}$, and clearly appears at about 74.5 K  at 19.5 GPa  }
		\label{Tc_diff.fig}	
\end{center}		
\end{figure}

\begin{figure}[]
\begin{center}	
	\includegraphics[scale=0.33]{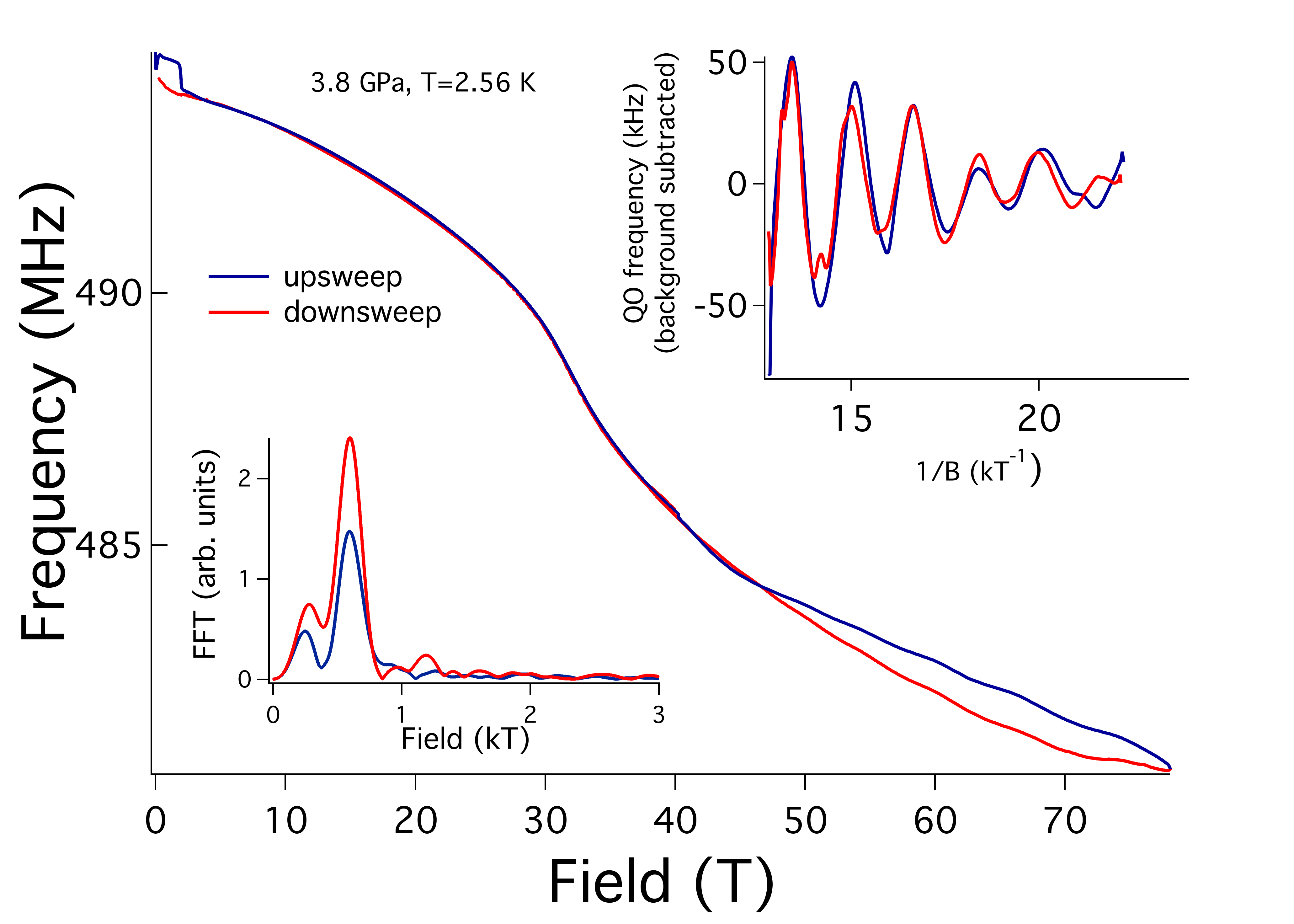}
		\caption{Main panel: TDO frequency vs field for a sample at 2.56 K and 3.8 GPa, taken in the 85 T duplex coil at HLD-EMFL. Top right panel: oscillatory part, background-subtracted vs 1/B.  Bottom left panel: Fourier transformation of the signal shown in the top right panel. }
		\label{SI2_Layout85T.fig}	
\end{center}		
\end{figure}

\begin{figure}[]
\begin{center}	
	\includegraphics[scale=1.3]{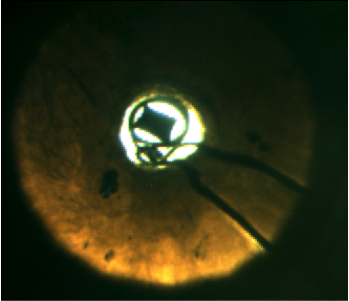}
		\caption{Image taken through  the diamond anvil of the 25 GPa cell,  which shows a typical TDO DAC set up - the rectangular sample can be seen in the coil which resides in the clear space of the gasket filled with 4:1::methanol:ethanol pressure medium with the coil leads coming out over the insulating insert of the gasket. The circular, semi-transparent area around the sample space is the diamond powder and epoxy gasket, and the darker region is the metallic ring confining it. A few ruby spheres can be distinguished between the top part of the coil and the gasket.}
		\label{SI3_photo2_setup.fig}	
\end{center}		
\end{figure}

\begin{figure}[]
\begin{center}	
	\includegraphics[scale=0.29]{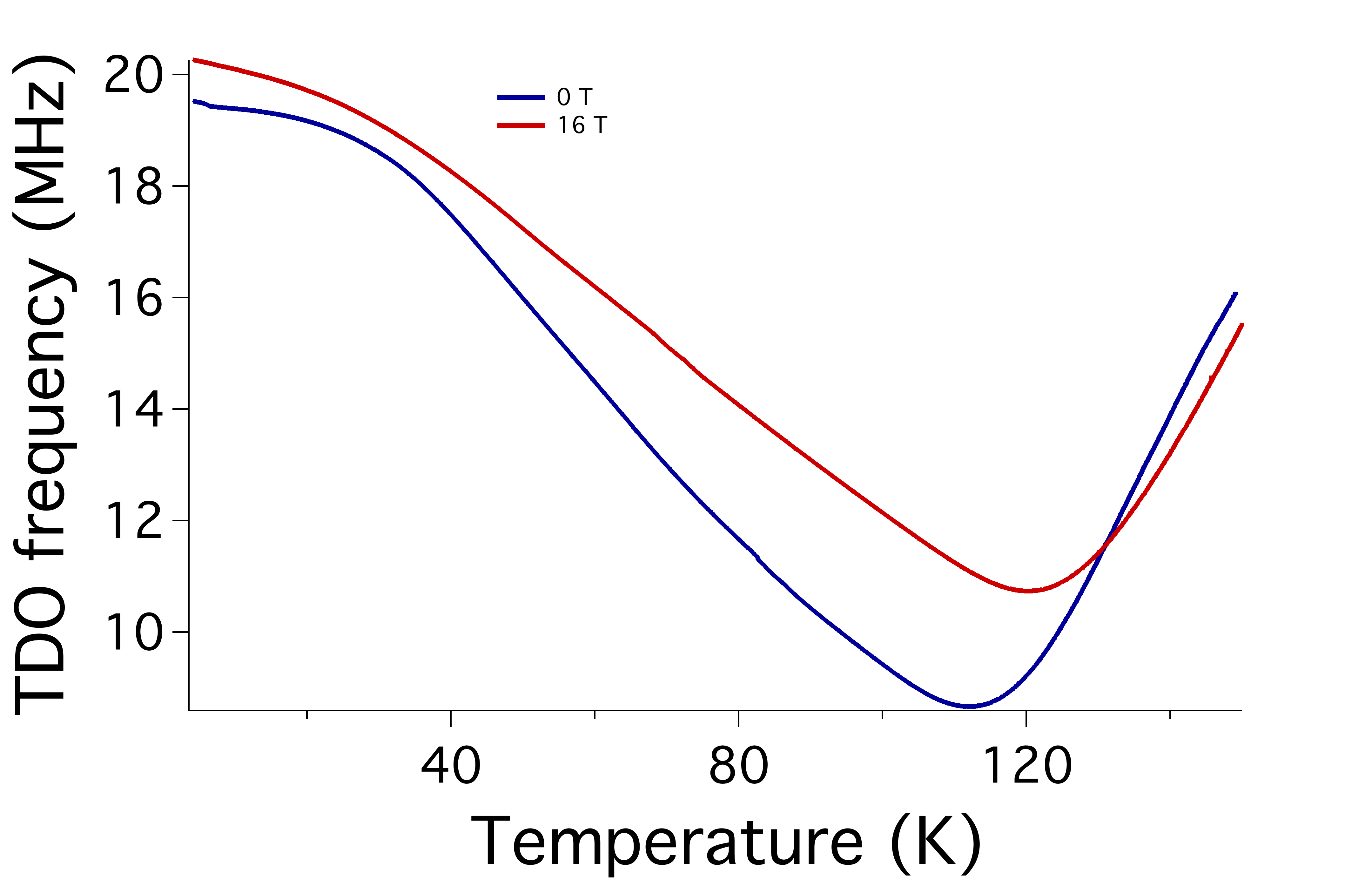}
		\caption{Evolution of the TDO frequency as a function of temperature for the 25 GPa cell,  at 0 and 16 T.  The broad feature observed in the 0 T trace cannot be interpreted as the superconducting transition widened by non-hydrostatic conditions as  an applied field of 16 T does not shift it to lower temperatures. }
		\label{SI4_Cell_25GPa.fig}	
\end{center}		
\end{figure}

\begin{figure}
\centering
\begin{subfigure}[]{}
\centering
\includegraphics[width=0.45\textwidth]{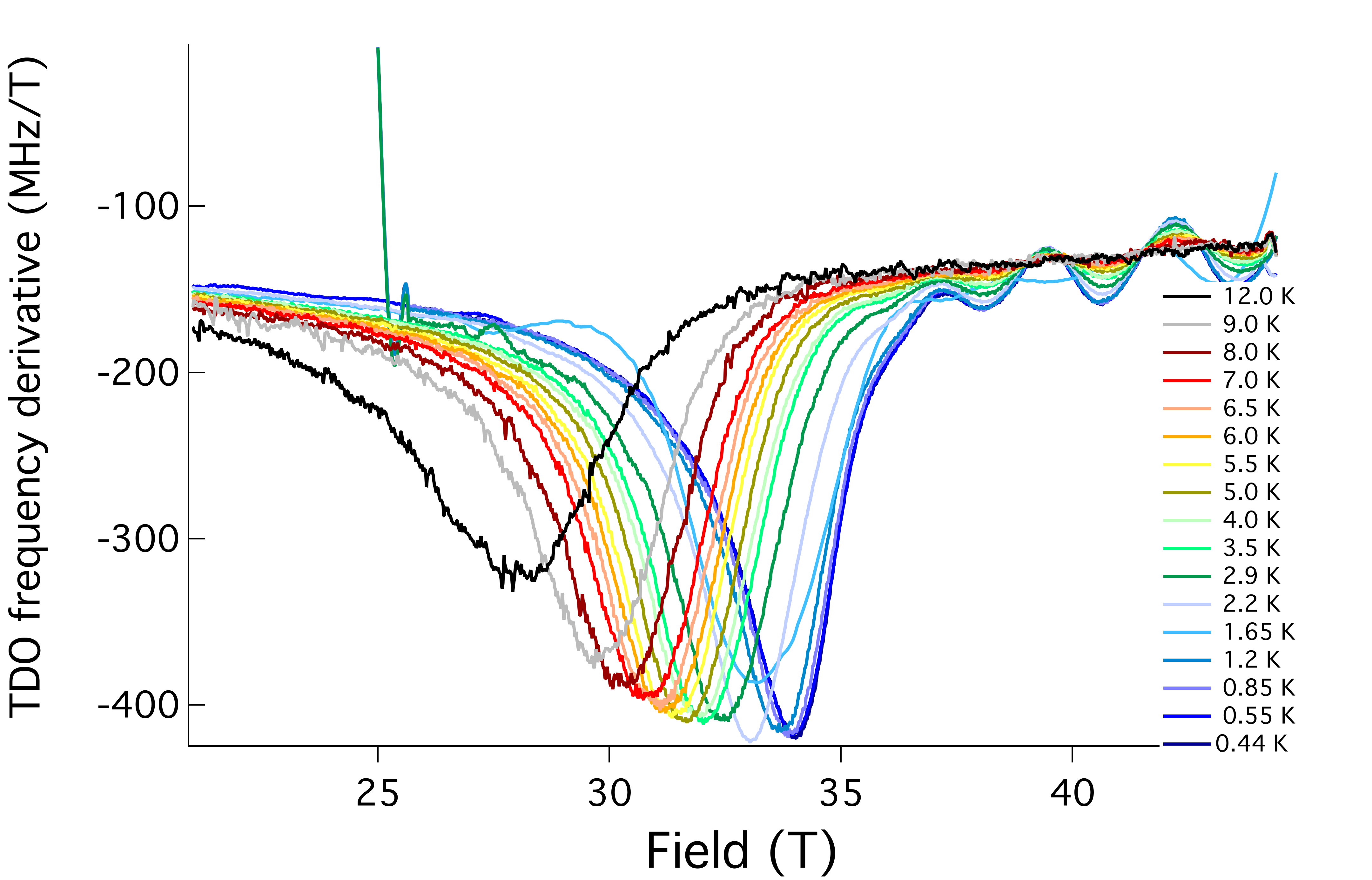} 
\end{subfigure}

\begin{subfigure}[]{}
\centering
\includegraphics[width=0.45\textwidth]{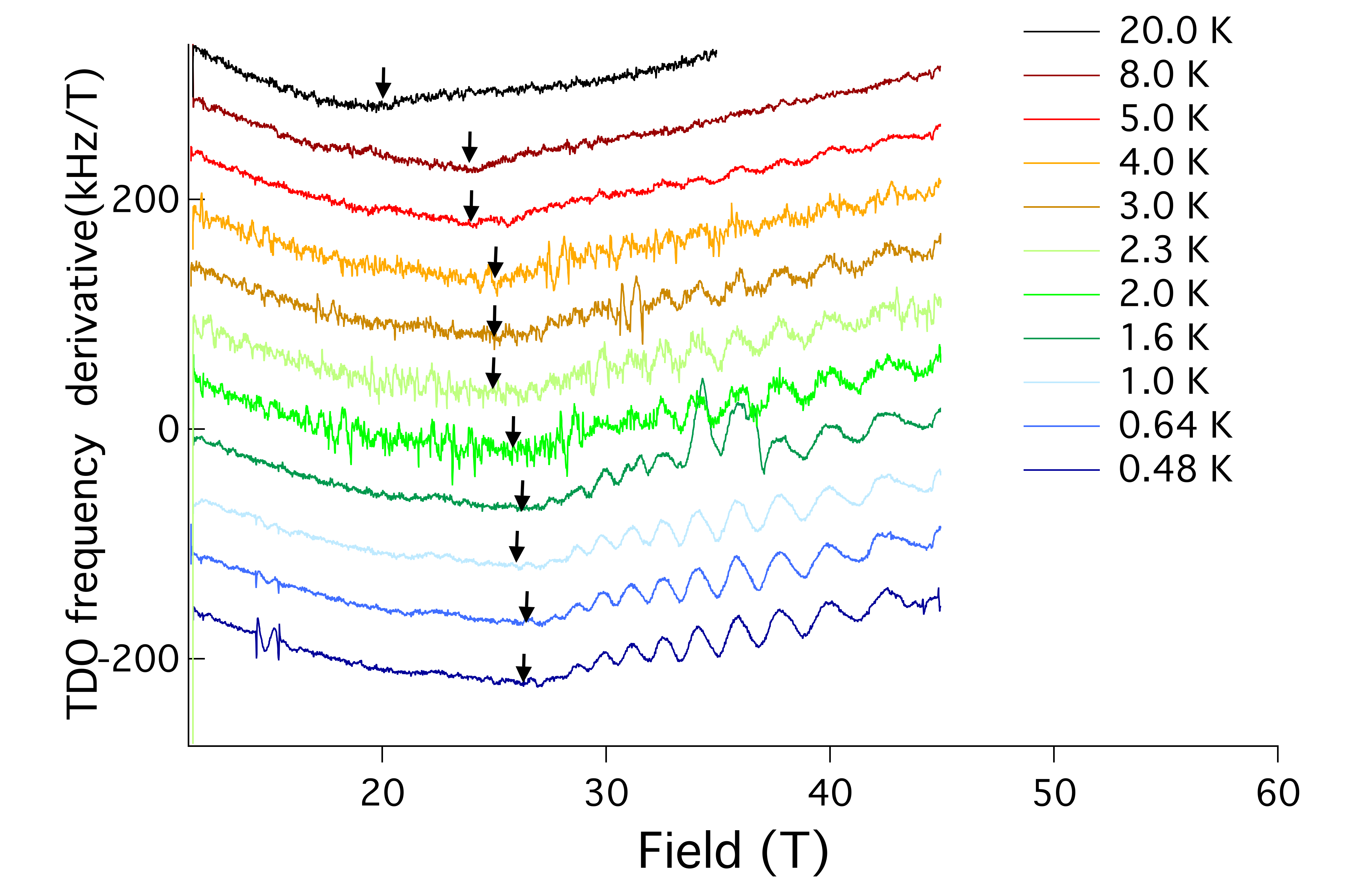} 
\end{subfigure}
 \caption{(a) derivative of the TDO frequency vs field at 4.7 GPa for various temperatures. The critical field is clearly identified as the minimum in the derivative,  followed by the onset of quantum oscillations. (b) derivative of the TDO frequency vs field at 18.8 GPa.  The minimum in the derivative,  compared with that obtained for 4.7 GPa, followed by the onset of quantum oscillations, is interpreted as the critical field at this pressure and is indicated by arrows.}
\label{Tdiff.fig}
\end{figure}

\begin{figure}[]
\begin{center}	
	\includegraphics[scale=0.29]{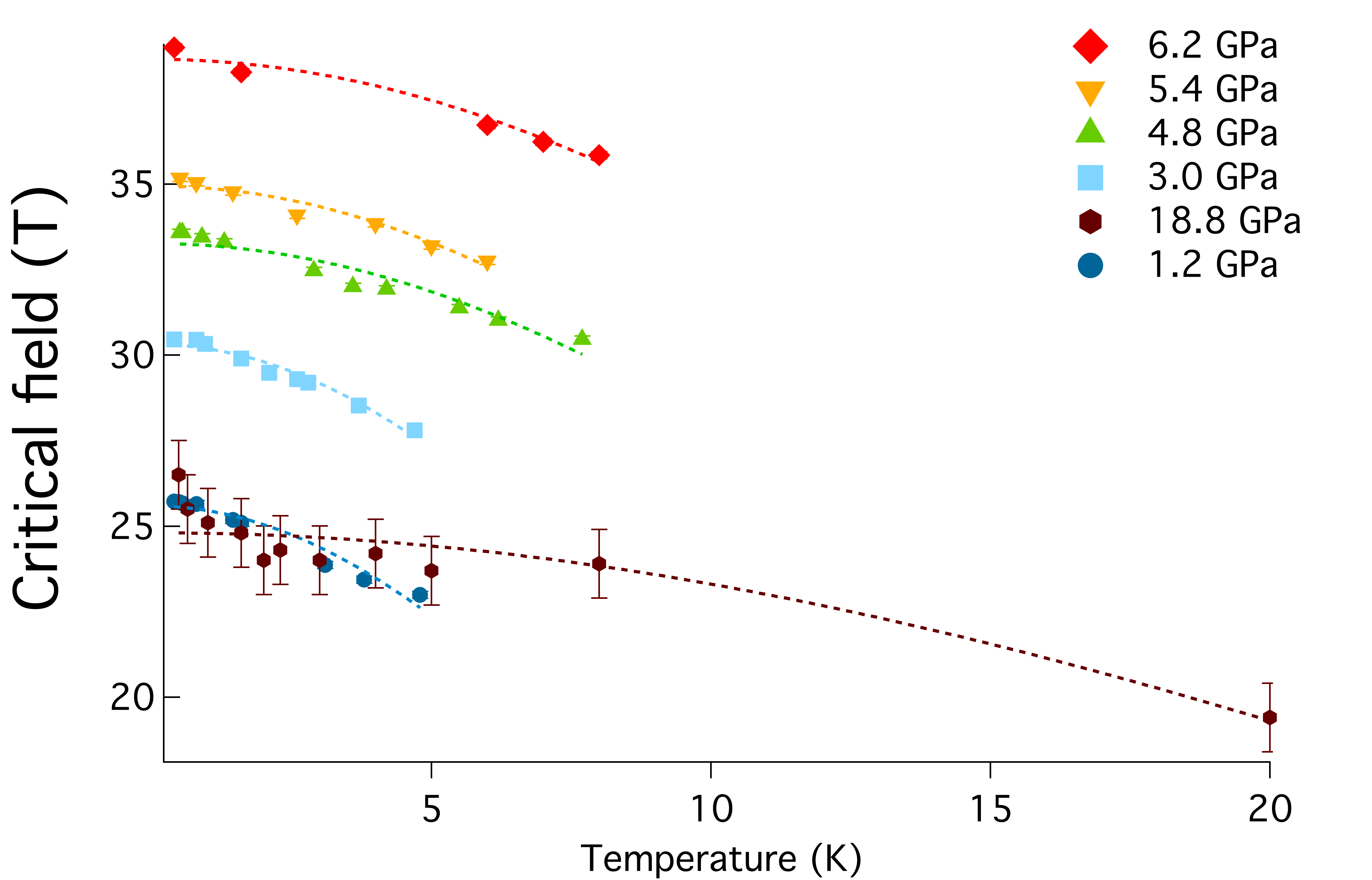}
		\caption{Evolution of the upper critical field as a function of temperature for several pressures.  The dotted lines are GL fits.  The minimum in the derivative defined as $H_{c2}$ observed at 28.4 T ( Fig. \ref{fig4}) for 18.8 GPa is plotted here along with data from the same feature found at lower pressures and seems to follow the similar trend.}
		\label{SI5_Hc_T_dep.fig}	
\end{center}		
\end{figure}

\begin{figure}[]
\begin{center}	
	\includegraphics[scale=0.29]{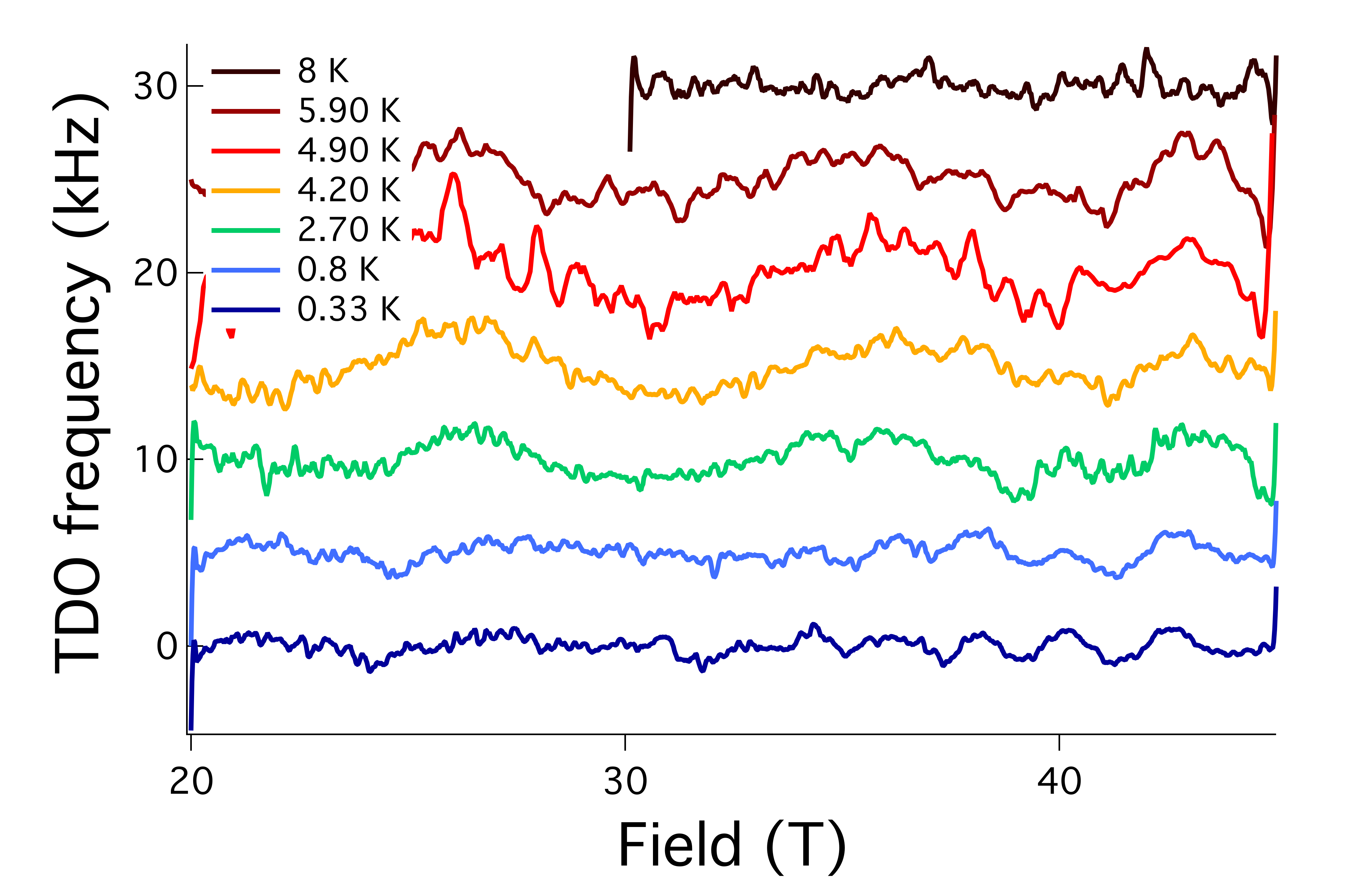}
		\caption{Background-subtracted TDO frequencies at 25 GPa,  showing the temperature evolution of the QO amplitudes, existing up to 2.7 K.   The traces are vertically shifted for clarity.}
		\label{SI6.fig}	
\end{center}		
\end{figure}

\end{document}